\documentclass[AMA,STIX1COL]{WileyNJD-v2}

\articletype{Research Article}%

\usepackage[english]{babel}
\usepackage{arydshln}

\usepackage{amsmath}
\usepackage{graphicx}
\usepackage{statmath}
\usepackage{amsthm}
\DeclareMathOperator{\logit}{logit}
\usepackage{float}
\usepackage{ulem}
\usepackage{enumitem}
\usepackage{graphicx}
\usepackage{hyperref}
\usepackage{caption}
\usepackage{setspace}
\usepackage{lscape}
\usepackage{booktabs}
\definecolor{ForestGreen}{RGB}{34,139,34}
\usepackage{multirow}
\usepackage{threeparttable}
\usepackage{algorithm}
\usepackage{algpseudocode}

\begin{document}

\title{Improving Variance and Confidence Interval Estimation in Small-Sample Propensity Score Analyses: Bootstrap vs. Asymptotic Methods}

\author[1,2]{Baoshan Zhang}

\author[1,2]{Sean M. O'Brien}

\author[1]{Yuan Wu}

\author[1,2]{Laine E. Thomas*}

\authormark{Baoshan Zhang \textsc{et al}}

\address[1]{\orgdiv{Department of Biostatistics and Bioinformatics}, \orgname{Duke University}, \orgaddress{\city{Durham}, \state{North Carolina}, \country{USA}}}
\address[2]{\orgdiv{Duke Clinical Research Institute} \orgname{Duke University Medical Center}, \orgaddress{\city{Durham}, \state{North Carolina}, \country{USA}}}

\corres{*\email{laine.thomas@duke.edu}}

\presentaddress{2424 Erwin Rd, Durham, NC 27705}

\abstract{
Propensity score (PS) methods are widely used to estimate treatment effects in non-randomized studies. Variance is typically estimated using sandwich or bootstrap methods, which can either treat the PS as estimated or fixed. The latter is thought to be conservative. Comparisons between the sandwich and bootstrap estimators have been compared in moderate to large sample sizes, favoring the bootstrap estimator. With the growing interest in treatments for rare disease and externally controlled clinical trials, very small sample sizes are not uncommon and the asymptotic properties of sandwich estimators may not hold.  Bootstrap methods that allow for PS re-estimation can also generate problems with quasi-separation in small samples. It is unclear whether it is safe to prefer sandwich estimators or to assume that treating the PS as fixed is conservative. We conducted a Monte Carlo simulation to compare the performance of bootstrap versus sandwich variance and CI estimators for average treatment effects estimated with PS methods. We systematically evaluated the impact of treating the PS as fixed versus re-estimating it. These methodological comparisons were performed using Inverse Probability of Treatment Weighting (IPTW) and Augmented Inverse Probability of Treatment Weighting (AIPW) estimators. Simulations assessed performance under various conditions, including small sample sizes and different outcome and treatment prevalences. We illustrate the differences in our motivating example, the LIMIT-JIA trial. We show that the sandwich estimators can perform quite poorly in small samples, and fixed PS methods are not necessarily conservative. A stratified bootstrap avoids quasi-separation and performs well. Differences were large enough to alter statistical conclusions in our motivating example, LIMIT-JIA.
}
\keywords{inverse probability of treatment weighting, propensity score, bootstrap, variance estimation}

\maketitle

\section{Introduction}
 
Propensity score (PS) methods are widely used in observational studies to adjust for confounding and estimate treatment effects. The PS is the probability of a subject being treated based on the participants' observed or measured baseline covariates (Rosenbaum \& Rubin, 1983 \cite{rosenbaum1983central}). One common approach to PS analysis is inverse probability of treatment weighting (IPTW), where each subject is assigned a weight equal to the inverse of their probability of receiving the treatment they received. Other PS weights can be derived similarly for a range of estimands. Among them, overlap weights (OW) targets those subjects for whom there is the greatest equipoise in treatment selection, and minimizes problems like extreme weights and chance imbalance (Li et al., 2018 \cite{li2018balancing};  Li et al., 2019 \cite{li2019addressing}).  

Asymptotic variance or sandwich estimators are well developed and widely used for estimators using IPTW and OW (Lunceford \& Davidian, 2004 \cite{lunceford2004stratification}; Li et al. 2019 \cite{li2019addressing}). However, these asymptotic estimators may perform poorly in small sample sizes, where bootstrap resampling methods provide an alternative (Austin, 2022 \cite{austin2022bootstrap}; Efron \& Tibshirani, 1994 \cite{efron1994introduction}).  Both types of estimators involve the PS and may regard the PS as fixed (and known) or estimated. Treating the PS as fixed is generally expected to yield conservative or overestimated sandwich variance, although this expectation is based on large-sample asymptotics (Lunceford \& Davidian, 2004 \cite{lunceford2004stratification}; Abadie \& Imbens, 2016 \cite{abadie2016matching}; Reifeis \& Hudgens, 2022 \cite{reifeis2022variance}; Mao et al., 2019 \cite{mao2019propensity}; Kostouraki et al. \cite{kostouraki2024variance}), which is not guaranteed to closely approximate finite sample properties. Moreover,  Reifeis \& Hudgens (2022) \cite{reifeis2022variance} showed that treating the PS as fixed can produce anti-conservative results when targeting the average treatment effect on the treated (ATT) estimand, even in large samples. This finding underscores that the common assumption of conservatism does not hold uniformly across estimands or sample sizes, motivating further evaluation across a range of study contexts.

Analogously, bootstrap methods for variance estimation typically treat the PS as estimated and re-fit the propensity model within each bootstrap sample.\cite{austin2022bootstrap}  However, this approach can be computationally intensive, particularly for complex PS models, and it remains unclear whether treating the PS as fixed within the boostrap process yields conservative inferences in small samples.

The preceding work has focused on total sample sizes of n>1000 \cite{mao2019propensity}, n>500 \cite{reifeis2022variance}, and n>250 \cite{austin2022bootstrap}. However, the growing use of PS methods for rare diseases\cite{narayanaswami2024comparative} and hybrid clinical trials\cite{ClinicalTrials_NCT03841357, samad2015management} (or single arm trials with external controls), increases the importance of understanding very small sample sizes. In particular, our motivating example is the LIMIT-JIA clinical trial of abatacept for juvenile idiopathic arthritis. Part II of LIMIT-JIA was a single-arm trial that included 76 patients who received abatacept, compared to 596 controls from the CARRA Registry, who did not begin abatacept as an initial treatment. In addition to the primary comparison, pre-specified subgroup analysis was a primary study aim of LIMIT-JIA, introducing even smaller sample sizes. In these small samples, excessive conservative is undesirable because precision is crucial, yet type I error must also be maintained. Beyond this motivating example, small samples are not rare in medical research. Among the 176 medical studies employing PS methods reporting their sample sizes reviewed by Stürmer et al. (2006) \cite{sturmer2006review}, the 25th percentile was 687 subjects, and approximately one-third (33\%) had sample sizes below 1,000 \cite{austin2022bootstrap}. These findings highlight the frequent occurrence and practical relevance of evaluating PS methods specifically in small-sample contexts.

In addition to variance estimation, ATE confidence interval estimation is also of vital importance to explore. To our knowleadge, previous studies have primarily considered standard normal (Wald) CIs for IPTW estimators(Austin, 2022\cite{austin2022bootstrap}; Reifes \& Hudgens, 2022 \cite{reifeis2022variance}).  Austin (2022) \cite{austin2022bootstrap} did not include percentile-based bootstrap CI estimation due to its increased computational demands. Our study addresses this gap by comparing the performance of the standard normal CI with these computationally intensive bootstrap methods, namely the percentile, bias-corrected, and bias-corrected and accelerated intervals \cite{efron1994introduction}.

To enhance statistical efficiency and potentially achieve double robustness, augmented inverse probability of treatment weighting (AIPW) estimators incorporate regression models for the outcome variable alongside PS models (Bang \& Robins, 2005 \cite{bang2005doubly}; Mao et al., 2019 \cite{mao2019propensity}). Mao et al. (2019)\cite{mao2019propensity} proposed sandwich variance estimation equations for AIPW estimators using a family of balancing weights (Li et al., 2018 \cite{li2018balancing}) that account for uncertainty in both the PS and outcome models. They evaluated these estimators primarily in large samples. More recently, Matsouaka et al. (2023) \cite{matsouaka2023variance} developed sandwich-based variance estimators for augmented ATT and ATC estimators and compared them with bootstrap approaches, considering model misspecification. However, to our knowledge, no study has directly compared sandwich- and bootstrap-based variance and CI estimation methods for AIPW estimators based on standard IPTW or OW, especially in small samples.

This study addresses these gaps by evaluating variance and CI estimation methods under small sample sizes (e.g., $n < 250$). 
We also specifically investigate how outcome prevalence and treatment prevalence influence performance. 
Section \ref{Sec: EstMethods} reviews existing statistical methods for variance and CI estimation for the ATE, covering various sandwich and bootstrap procedures. 
This section also describes an empirical, fully data-based version of the sandwich estimator that avoids substituting model-based expectation. Furthermore, methods relevant to the Average Treatment Effect on the Overlap Population (ATO) and augmented Inverse Probability Weighting (AIPW) estimators are discussed in this section.
Details of the Monte Carlo simulation design, including the factors varied, the data generating processes, and the metrics for analyzing simulation outcomes, are presented in Section \ref{Sec: MCmethod}. 
Section \ref{Sec: MCresults} presents the comparative results derived from these simulations, accompanied by detailed commentary. 
An application illustrating the performance of these variance and CI estimators using data from a real-world study (LIMIT-JIA) is provided in Section \ref{Sec: RealAna}. 
Finally, Section \ref{Sec: Discussion} summarizes the key findings, offers practical recommendations, acknowledges limitations of the study, and outlines potential directions for future work.

\section{Notation and Methods} \label{Sec: EstMethods}
In this study, we denote $Z$ as the binary treatment indicator, where $Z=1$ indicates assignment to the treatment group and $Z=0$ to the control group. Let $Y$ be a binary outcome, and let $\mathbf{X} = (X_1, \ldots, X_p)$ denote a vector of $p$ baseline covariates. Thus, the observed data are represented as $O = \{(Y_i, \mathbf{X}_i, Z_i), i = 1, \ldots, n\}$, corresponding to a sample of $n$ independent subjects drawn from a target population, often referred to as a super-population in causal inference.

We adopt the potential outcomes framework of Imbens \& Rubin (2015) \cite{imbens2015causal}, wherein each individual has two potential outcomes, $Y(0)$ and $Y(1)$, corresponding to the outcomes that would be observed under control and treatment, respectively. To ensure this treatment effect is identifiable from the observed data, we rely on three key assumptions (Rosenbaum \& Rubin, 1983)\cite{rosenbaum1983central}: (1) Stable Unit Treatment Value Assumption (SUTVA) which posits that an individual's outcomes are unaffected by the treatment status of others; (2) Unconfoundedness, which states that treatment assignment is independent of potential outcomes conditional on the observed covariates (i.e., $(Y(0), Y(1) \perp \!\!\! \perp Z \mid \mathbf{X})$); (3) Positivity, which requires that all individuals have a non-zero probability of receiving either treatment.  The conditional average treatment
effect (CATE) is $\Delta_i := \E[Y_i(1)-Y_i(0)\mid \mathbf{X}_i]$. Under the balancing weight framework, weighted average treatment effect estimands are defined as: 
\begin{equation}\label{Eq: Estimand}
    \Delta_\omega = \frac{\E[\omega(e_i)\Delta_i]}{\E[\omega(e_i)]},
\end{equation}
where the expectation is taken over the sampling distribution and $\omega(.)$ is a function of the propensity score (PS), $e(\mathbf{X}_i):= P(Z_i=1 \mid \mathbf{X}_i)$, which is the true conditional probability of receiving treatment given baseline covaraites. Different choices of $\omega(.)$ can target different populations of interests.

In practice, the true PS is unknown and typically estimated via logistic regression of $Z$ on $\mathbf{X}$, yielding
\[
  e_i = e(\mathbf{X}_i, \beta) = \frac{\exp(\mathbf{X}_i^\top \beta)}{1 + \exp(\mathbf{X}_i^\top \beta)},
\]
where $\beta$ is the vector of regression coefficients. Although various flexible approaches—such as machine learning methods including decision trees (Lee et al., 2010)\cite{lee2010improving} and the Super Learner (Pirracchio et al., 2015 )\cite{pirracchio2015improving}—can be used to estimate PS, in low-dimensional settings, logistic regression remains a practical and robust choice. Denote the estimated PS vis logistic regresion by $\hat{e}_i = e(\mathbf{X}_i, \hat{\beta})$. The balancing weight \cite{li2018balancing, hirano2003efficient} is defined as
\[
  W_i = \frac{\omega(\hat{e}_i)}{Z_i \hat{e}_i + (1 - Z_i)(1 - \hat{e}_i)}.
\]
With this weighting scheme, a consistent estimator for the estimand in formula (\ref{Eq: Estimand}) via balancing weight is given by
\begin{equation}\label{eq: WATE}
      \hat{\Delta}_\omega = \frac{\sum_{i=1}^n Z_i Y_i W_i}{\sum_{i=1}^n Z_i W_i} - \frac{\sum_{i=1}^n (1 - Z_i) Y_i W_i}{\sum_{i=1}^n (1 - Z_i) W_i}.
\end{equation}
This flexible framework allows us to target different populations of interest by specifying the functional form of $\omega(e_i)$. A widely used estimator for the ATE proposed by Lunceford \& Davidian (2004) \cite{lunceford2004stratification} corresponds to the choice $\omega(e_i) \equiv 1.$
% , resulting in:
% \begin{equation}\label{eq: Point_ATE}
%   \widehat{\Delta}_{\text{ATE}} = \left( \sum_{i=1}^n \frac{Z_i}{\hat{e}_i} \right)^{-1} \sum_{i=1}^n \frac{Z_i Y_i}{\hat{e}_i} - \left( \sum_{i=1}^n \frac{1 - Z_i}{1 - \hat{e}_i} \right)^{-1} \sum_{i=1}^n \frac{(1 - Z_i) Y_i}{1 - \hat{e}_i}.
% \end{equation}
% Stabilized weights can improve variance efficiency and yield more stable point estimates (Hirano et al., 2003\cite{hirano2003efficient}), making this estimator our preferred approach for estimating the ATE in this study.
Additionally, Li et al. (2018) \cite{li2018balancing} proposed a causal estimand targeting the average treatment effect for the overlap population (ATO), defined by setting $\omega(e_i) = e_i(1 - e_i)$. This estimand focuses on the subpopulation with the greatest clinical equipoise. 
% The corresponding estimator is given by
% \begin{equation}\label{eq: Point_ATO}
%   \widehat{\Delta}_{\text{ATO}} = \frac{\sum_{i=1}^n Z_i Y_i (1 - \hat{e}_i)}{\sum_{i=1}^n Z_i (1 - \hat{e}_i)} - \frac{\sum_{i=1}^n (1 - Z_i) Y_i \hat{e}_i}{\sum_{i=1}^n (1 - Z_i) \hat{e}_i}.
% \end{equation}

\subsection{Statistical Analyzes of Variance Estimation for ATE} \label{Sec: VarEst}

% Estimand of average treatment causal effect is defined as
% \[\Delta = E[Y(1)-Y(0)]\]
% and estimation of $\Delta_{\text{ATE}}$ is thus of central interest in comparing treatments. 

The estimation of variance for ATE is critical for constructing valid confidence intervals and conducting hypothesis tests. A key consideration is whether to treat the coefficients of the PS model, $\beta$, as fixed (known) or as estimated from the data. This choice leads to different large-sample theoretical variances. When the PS is treated as known, the large-sample variance of the ATE estimator is given by
\[
\Sigma^*_\text{ATE} =\E\left\{
\frac{[Y(1)-\mu_1]^2}{e} + \frac{[Y(0)-\mu_0]^2}{1-e}\right\},
\]
where $\mu_1 = \E[Y(1)]$ and $\mu_0 = \E[Y(0)]$. However, when the PS model is estimated, the additional constraint on the data reduces the sampling variability. The resulting large-sample variance is
\[
\Sigma_\text{ATE}= \Sigma_\text{ATE}^* - \mathbf{H}_{\beta}^\top\mathbf{E}_{\beta\beta}^{-1}\mathbf{H}_{\beta}, \quad \text{where }~ \mathbf{H}_{\beta} = E\left[\left(\frac{Y(1)-\mu_1}{e}+\frac{Y(0)-\mu_0}{1-e}\right)e_\beta\right],
\]
and $e_\beta =\partial/\partial\mathbf{\beta}\{e(\mathbf{X,\beta})\}.$ As shown by Lunceford \& Davidian (2004) \cite{lunceford2004stratification}, this demonstrates the theoretical benefit of using an estimated PS in large samples, as it yields a smaller asymptotic variance. While this theory provides valuable context, practical application requires estimators that can be calculated from sample data. Below, we introduce and compare several such estimators.

\subsubsection{Asymptotic Variance Estimators}\label{Sec: VarEst_Sand}

In this section, we derive variance estimators for the ATE, which, as previously noted, is a special case of the WATE estimator in Equation~(\ref{eq: WATE}) corresponding to the choice $\omega(e_i) \equiv 1$. For this specific case, the point estimator simplifies to the standard IPTW estimator, $\hat{\Delta}_\text{ATE} = \hat{\mu}_1 - \hat{\mu}_0$, where
$$
\hat{\mu}_{1}=\left( \sum_{i=1}^n \frac{Z_i}{\hat{e}_i}\right)^{-1} \sum_{i=1}^n \frac{Z_iY_i}{\hat{e}_i} \quad \text{and} \quad \hat{\mu}_{0}=\left( \sum_{i=1}^n \frac{1-Z_i}{1-\hat{e}_i} \right)^{-1} \sum_{i=1}^n \frac{(1-Z_i)Y_i}{1-\hat{e}_i}.
$$

\paragraph{Case 1: Estimators Accounting for Propensity Score Estimation}

We first consider estimators that account for the efficacy introduced by estimating the PS model coefficients $\beta$. The sandwich estimator from Lunceford \& Davidian (2004) \cite{lunceford2004stratification} is a standard approach derived from M-estimation theory \cite{stefanski2002calculus}. From this perspective, their estimator can be interpreted as a version where the component of the M-estimation derivative matrix corresponding to the outcome means ($\mu_0, \mu_1$) is simplified by replacing it with its theoretical expectation (the identity matrix). This simplification is valid under the crucial assumption that PS model is correctly specified. This model-based sandwich (MS) estimator has the following form:
\begin{equation}
    \hat{\Sigma}_\text{MS} =
    n^{-1}\sum_{i=1}^n\left\{ \frac{Z_i(Y_i-\hat{\mu}_1)}{\hat{e}_i} - \frac{(1-Z_i)(Y_i-\hat{\mu}_0)}{1-\hat{e}_i}-(Z_i-\hat{e}_i)\hat{\mathbf{H}}_{1,\beta}^\top\hat{\mathbf{E}}_{\beta\beta}^{-1}\mathbf{X}_i
\right\}^2,
\end{equation}
where $\hat{\mu}_{1}=\left(
\sum_{i=1}^nZ_i/\hat{e}_i\right)^{-1}\sum_{i=1}^nZ_iY_i/\hat{e}_i$, $\hat{\mathbf{E}}_{\beta\beta} = n^{-1}\sum_{i=1}^n\hat{e}_i(1-\hat{e}_i)\mathbf{X}_i\mathbf{X}_i^\top$
\[
\hat{\mathbf{H}}_{1,\beta} = n^{-1}\sum_{i=1}^n \left\{\frac{Z_i(Y_i-\hat{\mu}_1)(1-\hat{e}_i)}{\hat{e}_i} +\frac{(1-Z_i)(Y_i-\hat{\mu}_0)\hat{e}_i}{1-\hat{e}_i}\right\}\mathbf{X}_i.
\]

In contrast, we employ a fully empirical form of the sandwich estimator (PES) that avoids susbtituting empirical quantities with their model-based expectations. It retains all empirical components of the variance calculation, ensuring robustness even if the PS model is mis-specified (the full derivation from the stacked estimating equations is provided in the Appendix). The defining feature of the PES is its commitment to a single, consistent weighting scheme derived directly from the M-estimation framework. Specifically, the inverse-sum-of-weights normalization factors---e.g., $(\frac{1}{n}\sum_{j} Z_j / \hat{e}_j)^{-1}$---are applied uniformly throughout the estimator's construction, appearing in its main terms and, crucially for internal consistency, also within the correction term, $\hat{\mathbf{H}}_{2,\beta}$. This ensures all components of the variance estimator follow a unified weighting scheme, which may offer enhanced robustness.
\begin{equation}
    \begin{aligned}
    \hat{\Sigma}_\text{PES} =
    n^{-1}\sum_{i=1}^n\left\{\left(\frac{1}{n}\sum_{j}\frac{Z_j}{\hat{e}_j}\right)^{-1} \frac{Z_i(Y_i-\hat{\mu}_1)}{\hat{e}_i} - \left(\frac{1}{n}\sum_{j}\frac{1-Z_j}{1-\hat{e}_j}\right)^{-1} \frac{(1-Z_i)(Y_i-\hat{\mu}_0)}{1-\hat{e}_i}-(Z_i-\hat{e}_i)\hat{\mathbf{H}}_{2,\beta}^\top\hat{\mathbf{E}}_{\beta\beta}^{-1}\mathbf{X}_i
    \right\}^2,
    \end{aligned}
\end{equation}
where
\[
\hat{\mathbf{H}}_{2,\beta} = n^{-1}\sum_{i=1}^n \left\{\left(\frac{1}{n}\sum_j\frac{Z_j}{\hat{e}_j}\right)^{-1}\frac{Z_i(Y_i-\hat{\mu}_1)(1-\hat{e}_i)}{\hat{e}_i} + \left(\frac{1}{n}\sum_j\frac{1-Z_j}{1-\hat{e}_j}\right)^{-1}\frac{(1-Z_i)(Y_i-\hat{\mu}_0)\hat{e}_i}{1-\hat{e}_i}\right\}\mathbf{X}_i.
\]

A third approach, which we denote as $\hat{\Sigma}_\text{NS}$, avoids deriving complex analytical formulas by leveraging numerical methods. This numerical sandwich (NS) strategy, uses techniques like numerical differentiation to approximate the required derivatives from the M-estimation framework, and is implemented in the widely used R package \texttt{PSweight} \cite{zhou2020psweight}. Unlike the analytical estimators \(\hat{\Sigma}_\text{MS}\) and \(\hat{\Sigma}_\text{PES}\), $\hat{\Sigma}_\text{NS}$ computes the same underlying sandwich variance without a closed-form solution. While all three methods originate from the same estimating equation, they differ in implementation: \(\hat{\Sigma}_\text{MS}\) is partially model-based analytical solution, \(\hat{\Sigma}_\text{PES}\) is fully empirical analytical solution, and $\hat{\Sigma}_\text{NS}$ uses direct numerical approximation of derivatives.

\paragraph{Case 2: Estimator Treating  Propensity Score as Fixed}
Because it is widely used, we also consider the simplified case where the estimated PS, $\hat{e}_i$, are treated as known, fixed quantities. This approach ignores the uncertainty in the estimated PS, which prevents it from correctly reflecting the known efficiency benefits of estimating the PS model. We refer to this as a fixed sandwich (FS) estimator derived as
\begin{equation}
    \begin{aligned}
        \hat{\Sigma}^*_\text{FS} =  n^{-1}\sum_{i=1}^n\left\{\left(\frac{1}{n}\sum_{j}\frac{Z_j}{\hat{e}_j}\right)^{-1} \frac{Z_i(Y_i-\hat{\mu}_1)}{\hat{e}_i} - \left(\frac{1}{n}\sum_{j}\frac{1-Z_j}{1-\hat{e}_j}\right)^{-1} \frac{(1-Z_i)(Y_i-\hat{\mu}_0)}{1-\hat{e}_i} \right\}^2.
    \end{aligned}
\end{equation}
This estimator, $\hat{\Sigma}^*_\text{FS}$, is also implemented in the \texttt{PSweight} package and can be obtained by specifying that the PS is treated as fixed. 

In summary, we have presented $ 4$ asymptotic estimators for the ATE variance, which differ in their theoretical assumptions and computational approach.

\subsubsection{Variance Estimation based on Bootstrapping} \label{Sec: VarEst_Bootstrap}
We consider several bootstrap-based variance estimators for ATE \cite{efron1994introduction}. To clarify the fundamental procedure, all bootstrap methods discussed here estimate the variance by first generating $B$ bootstrap resamples from the original sample. For each of the $B$ resamples, a single ATE point estimate is calculated. The final bootstrap standard error estimate is then computed as the sample standard deviation of these $B$ ATE point estimates.  Then, our bootstrap procedures vary along two key dimensions: how the PS model is handled across resamples and the resampling strategy itself.

As above, bootstrapping can regard the PS as fixed, after estimating the model once on the initial sample. This method is valued for its computational efficiency and can provide more stable variance estimates in small or sparse samples where iterative re-estimation may be unreliable. However, it fails to account for finite-sample efficiency achieved by fitting the PS model to the data at hand\cite{mao2019propensity, kostouraki2024variance, hill2006interval}. In contrast, the gold-standard approach is to re-estimate the PS model within each bootstrap resample. %We distinguish these approaches using the prefixes \texttt{Fixed} and \texttt{Est}, respectively.

The second dimension addresses the resampling strategy itself, which is critical for ensuring the numerical stability of the PS model estimation in each replicate. A standard bootstrap draws resamples with replacement from the full dataset. In small samples this process can, by chance, produce replicates with severe treatment group imbalances, leading to model fitting failures from quasi-separation. To mitigate this, a stratified bootstrap resamples separately from within the treated and control groups, preserving the original treatment proportion in every replicate. This technique enhances stability and is a recommended practice in PS analyses \cite{hill2006interval}. %We denote these methods with the suffixes \texttt{std.B-} and \texttt{strt.B-}, respectively.

\begin{figure}[h]
\centering\includegraphics[width=\textwidth]{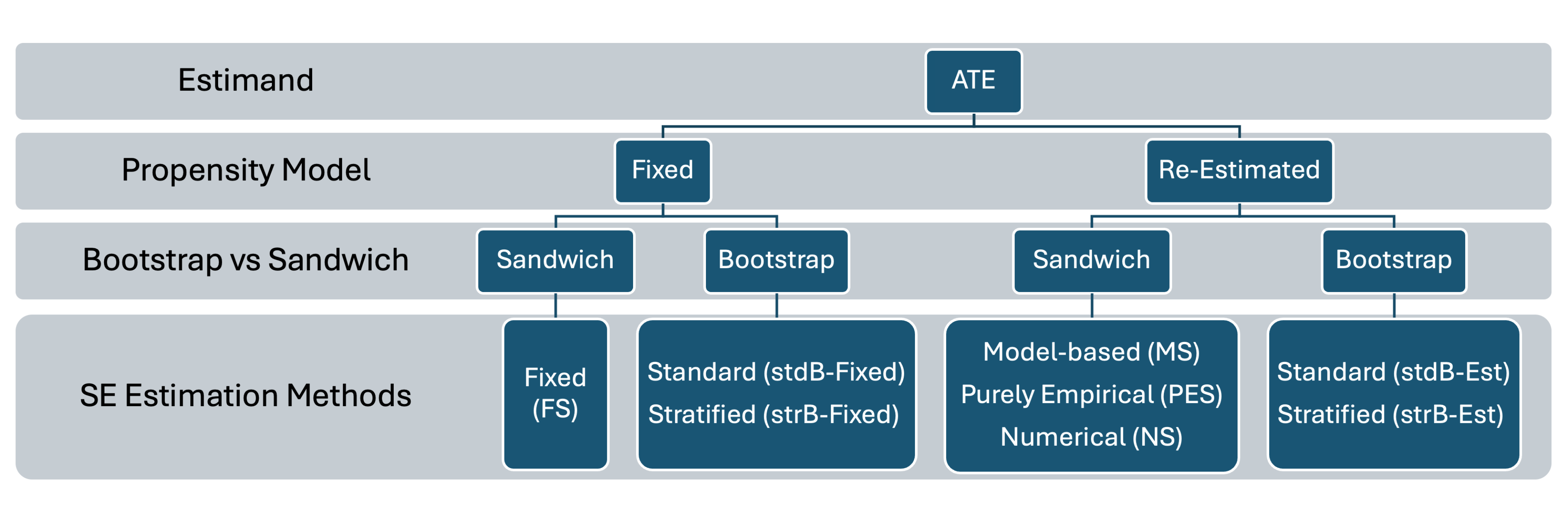}\captionsetup{ singlelinecheck=off} % Center the caption and allow for footnote
    \caption[Caption for LOF]{Flow Chart of Variance Estimation for ATE}
    \label{fig: FlowChart_Var}
\end{figure}

Combining these two dimensions yields four distinct bootstrap variance estimators for comparison. In summary, we present a hierarchy flow chart in Figure \ref{fig: FlowChart_Var} to show all mentioned methods (asymptotic and bootstrapping) we used to compare for the later simulation part.

\subsection{Statistical Analysis of Confidence Interval for ATE}\label{Sec: CI_ATE}

We generated bootstrap samples of the average treatment effect, sorted these samples in ascending order, and constructed 95\% confidence intervals (CIs) using the following methods:
\begin{enumerate}
    \item \textbf{Parametric Normal (Wald) Intervals}: 
    %(label as \texttt{Parmetric}) 
    These intervals utilized by Austin (2022) \cite{austin2022bootstrap}, use various standard error (SE) estimation methods detailed in subsection \ref{Sec: VarEst}. Specifically, the CIs are calculated as follows:
    \[ (\widehat{\Delta}_{\text{ATE}} -Z_{0.975}*\widehat{\text{SE}}_i, \quad \widehat{\Delta}_{\text{ATE}} + Z_{0.975}*\widehat{\text{SE}}_i)
    \]
    where $\widehat{\Delta}_{\text{ATE}}$ denotes the point estimate of ATE, $Z_{0.975}$ denotes the 97.5th percentile for the standardized normal distribution and $\widehat{\text{SE}_i}$ denotes the various SE estimation methods detailed in subsection \ref{Sec: VarEst}, which includes both sandwich estimators and bootstrapping methods. 
    \item \textbf{Percentile (Empirical) Interval}: 
    %(label as \texttt{Percentile}): 
    This method utilizes the empirical quantiles of bootstrap estimates. Specifically, the interval is defined by:
    \[
    (Q_{0.025}, \quad Q_{0.975}),
    \]
    where \(Q_{0.025}\) and \(Q_{0.975}\) represent the 2.5\% and 97.5\% empirical quantiles of the boostrap estimates, respectively. In the PS weighting context, this method was applied by Zhou et al. (2020) \cite{zhou2020psweight}.
    \item \textbf{Double Bootstrap Interval}: 
    %(label as \texttt{Double}): 
    This method, proposed by Efron \& Tibshirani (1994) \cite{efron1994introduction}, adjusts the percentile interval to enhance accuracy. The interval is computed as:
    \[(2\times\widehat{\Delta}_{\text{ATE}}-Q_{0.975},\quad 2\times\widehat{\Delta}_{\text{ATE}}-Q_{0.025}).
    \]
    \item \textbf{Bias-Corrected and Accelerated Interval}\cite{efron1994introduction} 
    %(label as \texttt{BCa})
    : This method refines the bootstrap percentile approach by adjusting for both bias and skewness in the bootstrap distribution through acceleration and bias-correction factors. In our analysis, we compute the BCa interval using the \texttt{bca} function from the \texttt{coxed} package in R.
\end{enumerate}

\subsection{Statistical Analyzes of Variance Estimation for ATO}

We also evaluate variance estimation methods for the Average Treatment effect in the Overlap population (ATO). The ATO is an alternative estimand that enhances stability by down-weighting individuals with extreme PS, which in turn yields a point estimator with smaller asymptotic variance \cite{li2018balancing}. Prior work suggests that the variance estimators are also more stable \cite{austin2022bootstrap}. It is unknown whether this advantage be amplified or diminished in the very small sample sizes considered here.

Our evaluation of variance estimators for the ATO mirrors the framework established for the ATE as is shown in Figure \ref{fig: FlowChart_Var}, encompassing both asymptotic and bootstrap approaches. However, we consider a slightly reduced set of estimators including, $\hat{\Sigma}_\text{PES}$ whose analytical form for the ATO is provided by Li et al. (2019) \cite{li2019addressing}, and ATO versions of $\hat{\Sigma}_\text{NS}$ and $\hat{\Sigma}^*_\text{FS}$.  Both numerical estimators are implemented in the \texttt{PSweight} package \cite{zhou2020psweight}, obtained by specifying overlap weights. Additionally, we apply the same four bootstrap-based procedures detailed in Section~\ref{Sec: VarEst_Bootstrap}.  

\subsection{Augmented by Outcome Model} \label{Sec: Aug}
Lastly, we further investigate how augmenting the estimator with an outcome model influences variance and confidence interval estimation. For any balancing weight $\omega(\cdot)$, the estimator in (\ref{eq: WATE}) can be augmented with an outcome regression model to yield a double-robust estimator (Lunceford \& Davadian, 2004 \cite{lunceford2004stratification}; Mao et al., 2019 \cite{mao2019propensity}). Let $m_1(\mathbf{X}, \alpha_1)$ be the outcome model for the treated group and $m_0(\mathbf{X}, \alpha_0)$ be the outcome model for the control group. For binary outcome $Y$, we could fit a logistic regression model for $Y$ on $(Z, \mathbf{X})$ to fit the outcome model $m_z(\mathbf{X}, \alpha_z)$. Then, the augmented estimator proposed is
\begin{equation}\label{eq: AWATE}
    \begin{aligned}
    \widehat{\Delta}_\text{Aug} 
    =& \frac{\sum_{i=1}^n \omega(\hat{e}_i)\bigl[m_1(\mathbf{X}_i, \hat{\alpha}_1) - m_0(\mathbf{X}_i, \hat{\alpha}_0)\bigr]}{\sum_{i=1}^n \omega(\hat{e}_i)} \\
    &+ \frac{\sum_{i=1}^n Z_i \bigl[Y_i - m_1(\mathbf{X}_i, \hat{\alpha}_1)\bigr] W_i}{\sum_{i=1}^n Z_i W_i}
    - \frac{\sum_{i=1}^n (1 - Z_i)\bigl[Y_i - m_0(\mathbf{X}_i, \hat{\alpha}_0)\bigr] W_i}{\sum_{i=1}^n (1 - Z_i) W_i}.
    \end{aligned}
\end{equation}
If PS is correctly specified, $\widehat{\Delta}_\text{Aug}$ remains consistent for its corresponding estimand regardless of whether the outcome models are correctly specified. If both the PS and outcome models are correctly specified, $\widehat{\Delta}_\text{Aug}$ yields a smaller variance in large samples (local efficiency )\cite{matsouaka2023variance,robins1994estimation, tsiatis2006semiparametric}. However, both double robustness and local efficiency are asymptotic properties, and our study focuses on small-sample performance of the augmented estimator.

% When fit the outcome model, We consider two scenarios: augmentation by the true outcome model and by a mis-specified outcome model. For variance estimation of $\Delta_\text{aug}$, we use the sandwich estimator by specifying the \texttt{out.formula} argument in the \texttt{PSweight} function \cite{zhou2020psweight}. Variance estimation via bootstrap follows the procedure in Section~\ref{Sec: VarEst_Bootstrap}, and confidence interval construction is analogous to that in Section~\ref{Sec: CI_ATE}. We consider both a fixed PS and an estimated PS in our analyses.

For the augmented estimator, we followed the framework established in Figure \ref{fig: FlowChart_Var}. Our investigation of sandwich variance estimators relies exclusively on a numerical approach. This strategy is chosen because analytical, closed-form solutions are complex to derive and depend heavily on the specific forms of both PS models and outcome models, which makes them inflexible for general application. We evaluate this numerical sandwich estimator under two cases: first, when properly accounting for the estimation of all nuisance parameters (in both the PS and outcome models), and second, in a simplified scenario that treats the PS model as fixed. Both of these numerical sandwich estimators are implemented via the \texttt{PSweight} package by specifying the outcome model \cite{zhou2020psweight}. The bootstrap procedures for the augmented estimator also directly mirror those applied to the framework of ATE, as detailed in Section~\ref{Sec: VarEst_Bootstrap}. This also includes the four variations based on PS model handling and resampling strategy.

\section{Monte Carlo Simulation Setting} \label{Sec: MCmethod}
We conducted a series of Monte Carlo simulations to compare the performance of aforementioned variance and confidence interval estimators. To be noted, for each iteration of the Monte Carlo simulation, a random sample of size $n$ is drawn from the super-population without replacement. Given that the sample proportion of treated patients is known in the actual data analysis, as noted by Austin (2022) \cite{austin2022bootstrap}, a stratified sampling technique is employed. This sampling technique ensures that the proportion of treated patients within each random sample remains fixed, mirroring the known proportion of the treatment arm in the actual data.

\subsection{Data Generating Process}
We followed Austin's model-specification for data generation \cite{austin2022bootstrap}. We simulated a super-population of $N=1,000,000$ subjects. For each subject $i$ in this super-population, we first generate ten variables $(V_1, V_2,\ldots, V_{10})$ from a multivariate normal distribution with a mean of zero, a variance of 1, and a constant pairwise correlation of 0.2. These variables were used to create ten baseline covaraites: the first five$(X_1,\ldots, X_5)$ were set equal to  $(V_1, V_2, \ldots, V_5)$, while the second five$(X_6,\ldots, X_{10})$ were binary covariates created by dichotomizing $(V_6, V_7, \ldots, V_{10})$ at their 10th, 20th, 30th, 40th, and 50th percentiles, respectively.  

After establishing the covariates, a binary treatment variable $Z_i$ for each subject $i$ was generated using the following logistic model: 
\begin{align*}
    \logit(p_{i,{\text{treat}}}) =& \alpha_{0,\text{treat}} + \log(1.1) x_{1,i} + \log(1.2) x_{2,i} + \log(1.5) x_{3,i} + \log(1.75) x_{4,i} + \log(2) x_{5,i}  \\
    & + \log(1.25) x_{6,i}+ \log(1.5) x_{7,i} + \log(2) x_{8,i} + \log(0.8) x_{9,i} + \log(0.5) x_{10,i}. \tag{Treatment Model}
\end{align*}
For each subject $i$, $Z_i$ was generated from a Bernoulli distribution with probability of event $p_{i,{\text{treat}}}.$ The intercept for the treatment model $\alpha_{0,\text{treat}}$ was determined using a bisection approach to ensure the prevalence of treatment $\Pr(Z=1)$ in the super-population matched a desired target. In our study, we range the prevalence of treatment from 0.1 to 0.5 with an increment of 0.1,solving for the corresponding $\alpha_{0,\text{treat}}$ at each level. 

Next, potential outcomes,$Y_i(1)$ and $Y_i(0)$, were generated for each subject using a logistic regression model structured as follows:
\begin{align*}
    \logit(p_{i,{\text{outcome}}}) =& \alpha_{0,\text{outcome}} + \alpha_\text{treat}Z_i + \log(2) x_{1,i} + \log(1.75) x_{2,i} + \log(1.1) x_{3,i} + \log(1.5) x_{4,i} + \log(1.2) x_{5,i} \\
    &+ \log(2) x_{6,i} + \log(1.5) x_{7,i} + \log(1.1) x_{8,i} + \log(1.25) x_{9,i} + \log(2) x_{10,i}. \tag{Outcome Model}
\end{align*}
Specifically, for each subject $i$, the probabilty of $Y_i(1)=1$ was determined by setting $Z_i=1$ in the Outcome model, and the probabilty of $Y_i(1)=0$ was determined by setting $Z_i=0$ in the Outcome model. The model parameters were calibrated sequentially. First, the intercept $\alpha_{0,\text{outcome}}$ was calibrated using a bisection method to control the prevalence of the untreated outcome,$\Pr(Y(0)=1$), matched a target value. This target oreavalence was varied from 0.1 to 0.5 in increments of 0.1. Second, for each of these target prevalence levels, the treatment effect parameter $\alpha_\text{treat}$ was iteratively adjusted until the true Average Treatment Effect (ATE), in terms of risk difference, reached -0.02. The true ATE was computed across the entire super-population as $\text{ATE}= \frac{1}{N}\sum_{i=1}^N[Y_i(1)-Y_i(0)]$.

In parallel, the true ATO was also determined from the super-population. Folloing Austin (2022)\cite{austin2022bootstrap}, the true ATO was calculated by applying the generalized weighted estimator $\hat{\Delta}_\omega$ from Equation (\ref{eq: AWATE}) to the entire super-population ($N=1,000,000$), specifying the weight function as $\omega(e_i)=e_i(1-e_i)$ and using the true propensity scores $p_{i,\text{treat}}$ as $e_i$. 

\subsection{Factors Varied in the MC Simulation}
The Monte Carlo simulation varies several factors. The first factor is the random sample size drawn from the super-population, taking values of 100, 150, 200, 300, 500, 750, and 1000. The second factor is the treatment prevalence, denoted as \(\Pr(Z=1)\), which ranges from 0.1 to 0.5 in increments of 0.1. The third factor is the prevalence of the outcome in the absence of treatment, \(\Pr(Y(0)=1)\), which also varies from 0.1 to 0.5 in increments of 0.1. These factors are fully crossed, resulting in a factorial design that systematically explores their combined effects.

Ideally, there should be $7\times 5\times 5= 175$ different scenarios to explore. However, considering that small sample sizes combined with extreme treatment proportions can lead to quasi-separation issues when fitting the PS using logistic regression, we excluded scenarios that frequently resulted in quasi-separation. Specifically, we omitted scenarios with a sample size of 100 and 150 when the treatment proportion was set at 0.1, as well as scenarios with a sample size of 100 when the treatment proportion was set at 0.2. In summary, the factorial design for this MC simulation study considered a total of 160 different scenarios.

\subsection{Estimation and evaluation}
We estimated the risk difference for both ATE and ATO. Both estimators are instances of the general weighted estimator shown in formula \eqref{eq: WATE}, distinguished only by the application of their respective weights. For the augmented estimator, we consider two scenarios for the outcome logistic model: one that is correctly specified and another that is mis-specified. In the correctly specified model, all covariates $X_1, X_2, \ldots, X_{10}$ are included, while the mis-specified model retains only five: three continuous covariates ($X_1, X_2, X_3$) and two binary covariates ($X_6, X_{10}$). In each random sample, we re-estimate the outcome model and subsequently re-estimate it for every bootstrap sample, thereby excluding any fixed-outcome-model approach for simplicity. For standard error and confidence interval estimation of the augmented estimator, we follow the procedures outlined in Section \ref{Sec: CI_ATE}. 

Standard errors (SE) for the risk difference were obtained as outlined in Section \ref{Sec: VarEst}. Given our emphasis on small sample sizes, we used 1,000 bootstrap samples per iteration, which is substantially greater than the 200 samples employed by Austin (2022) \cite{austin2022bootstrap}. The boostrap SE estimator was computed as the standard deviation of the point estimates across the B=1,000 bootstrap resamples.  Construction of the 95\% confidence intervals for the risk difference followed the methods in Section~\ref{Sec: CI_ATE}, including both standard normal intervals and those based on the bootstrap. We then assessed whether each interval contained the true treatment effect.

This process of drawing a random sample and applying the above steps was repeated 10,000 times. In each iteration, we recorded the point estimate of the risk difference, the asymptotic standard error of the estimate (from both the sandwich and bootstrap approaches), and the 95\% confidence interval. We compared the mean estimated standard error (across the 10,000 iterations) with the empirical standard deviation of the 10,000 point estimates. When this ratio equals one, it indicates that the estimated standard error aligns with the sampling distribution's standard deviation of the treatment-effect estimator. In addition, we computed the empirical coverage rate, defined as the proportion of confidence intervals that contained the true treatment effect.

\section{Monte Carlo Simulation-Results} \label{Sec: MCresults}

\subsection{Comparison of Estimated SE based on ATE}\label{Sec: RstSE_ATE}

The Monte Carlo simulation results for all evaluated methods are available in the Web Supplementary Material. In the main manuscript, we focus on four representative methods to highlight key findings regarding their performance and common use in practice. Methods in the Web Supplementary Material perform comparable to or worse than these selected methods:
\begin{enumerate}
    \item \texttt{FS} : The sandwich variance estimator ($\hat{\Sigma}^*_\text{FS}$) treating the PS model as fixed.
    \item \texttt{stdB-Fixed}: The standard bootstrap method treating the PS model as fixed.
    \item \texttt{NS}: The numerical
        sandwich variance estimator ($\hat{\Sigma}_\text{NS}$)  accounting for PS model estimation.
    \item \texttt{stdB-Est}: The standard bootstrap method with PS model re-estimation.
\end{enumerate}

\begin{figure}[h]
    \centering    \includegraphics[width=1\textwidth]{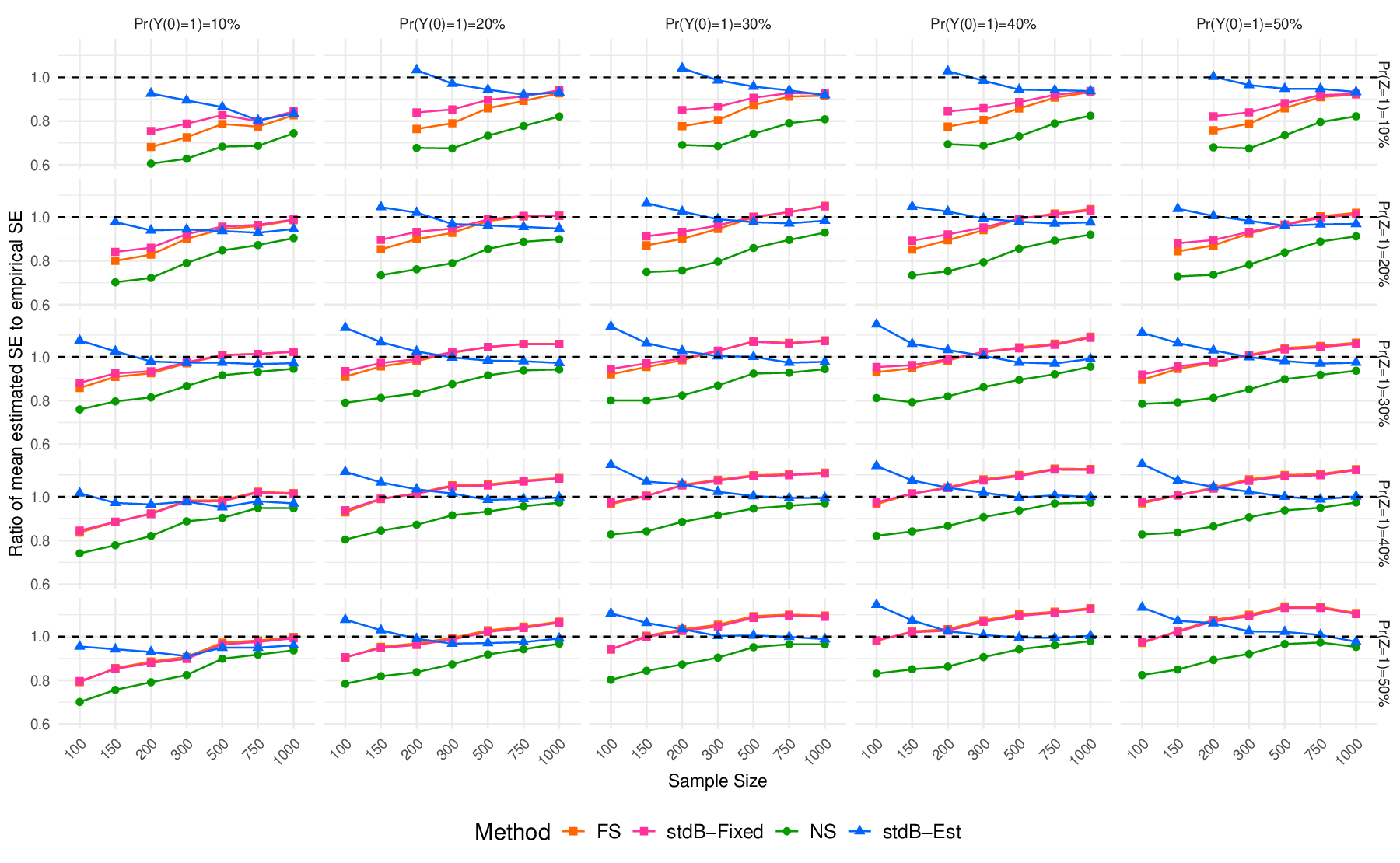}
    
    % Use \captionsetup for the main caption text
    \captionsetup{singlelinecheck=off} 
    
    \caption[Caption for LOF]{Ratio of Mean Estimated SE to Empirical SE for the ATE: Results for Selected Methods
        % --- Start of new "footnote" block ---
        \par % Add a line break
        \vspace{0.5ex} % Add a little vertical space
        \footnotesize 
        \par% Add another line break
        $^1$The standard errors (SEs) of the estimated SE for 10,000 runs of the Monte Carlo simulation are all less than 0.001 across all methods, demonstrating the precision of our estimated SE results.
        \par
     $^2$All sandwich variance estimations in this Figure were conducted using the \texttt{PSweight} function from the \texttt{PSweight} R package \cite{zhou2020psweight}.
    
        % --- End of new "footnote" block ---
    }
    \label{fig: SE.Ratio_Main}
\end{figure}

The ratio of the mean estimated SEs to the standard deviations of the estimated treatment effect for these four methods is reported in Figure \ref{fig: SE.Ratio_Main}. There are 25 panels in this Figure, one for each combination of setting of sample proportion of treatment $\Pr(Z=1)$ and prevalence of outcome if control $\Pr(Y(0)=1)$. Each panel consists of 4 lines, and each line stands for a selected method of SE estimation above. At each panel, the X-axis is the sample size for the random sample, and the Y-axis is the ratio of the mean estimated SE to the empirical SE. The black dashed horizontal line is where a ratio is equal to one. In addition, the SE of the estimated SE among all simulations is also compared and reported among these variance estimation methods in Web Supplementary Material to check the precision. 

% Based on Figure \ref{fig: SE.Ratio_Main} and the detailed results in the Web Supplementary Material, we identified several key findings.

A primary finding was that the choice between an asymptotic (sandwich) and a bootstrap approach had a varied impact, which depended critically on whether the PS model was treated as fixed or was re-estimated. 

When treating the PS as fixed, the difference between the bootstrap (\texttt{stdB-Fixed}) and sandwich (\texttt{FS}) estimators was relatively minor. Both methods exhibited a similar trend: the SE ratio was typically below one in small samples but increased with sample size, eventually exceeding one and becoming overly conservative. While this aligns with the theory that ignoring PS estimation is conservative in large samples \cite{lunceford2004stratification}, our findings showed this conservatism did not hold in small samples, where the trend was reversed. In sharp contrast, when accounting for PS model re-estimation, the difference between the two approaches was substantial. The two methods showed divergent behavior: the numerical sandwich estimator (\texttt{NS}) consistently produced the least conservative SE estimates (often underestimating variability in smaller samples), while the bootstrap estimator (\texttt{stdB-Est}) was uniformly the most conservative in small sample settings, with its ratio starting above one and converging towards it as the sample size grew.

Synthesizing these results, \texttt{stdB-Est} provides the most conservative and reliable variance estimates in small samples, while \texttt{NS} risks being the most anti-conservative.

\subsection{Comparsion of Estimated Confidence Intervals based on ATE}\label{Sec: RstCI_ATE}

The Monte Carlo results for the empirical coverage rates of 95\% CIs for all methods are reported in the Web Supplementary Material. In this subsection, we focus on five representative methods selected to illustrate key performance characteristics. Methods in the Web Supplementary Material perform comparably to these selcted methods. These methods are labeled by how the PS is handled and how the CI is constructed:
\begin{enumerate}
    \item \texttt{Par} \texttt{stdB-Fixed}: The parametric normal-theory CI constructed using the SE from \texttt{stdB-Fixed}.
    \item \texttt{Pct} \texttt{stdB-Fixed}: The CI formed directly from the empirical quantiles of the standard bootstrap distribution, treating the PS as fixed.
    \item \texttt{Par} \texttt{NS}: The parametric normal-theory CI constructed using the SE from  \texttt{NS}.
    \item \texttt{Par} \texttt{stdB-Est}: The parametric normal-theory CI constructed using the SE from \texttt{stdB-Est}.
    \item \texttt{Pct} \texttt{stdB-Est}: The CI formed directly from the empirical quantiles of the standard bootstrap distribution, with PS re-estimation in each replicate.
\end{enumerate}

\subsubsection{Results of Coverage Probability of Confidence Interval among selected Methods}
\begin{figure}[h]
    \centering
    \includegraphics[width=1\textwidth]{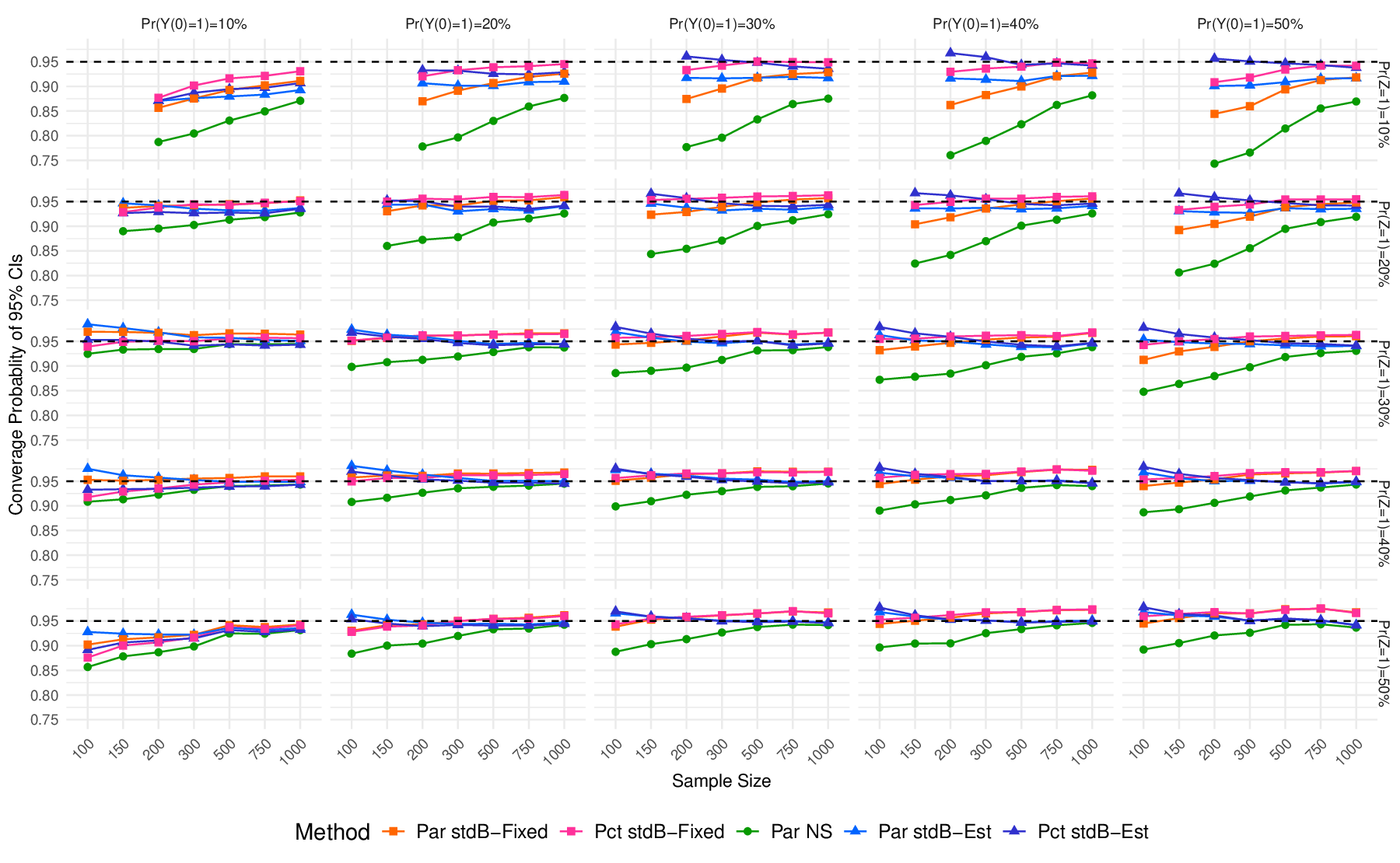}
    
    \captionsetup{singlelinecheck=off}
    
    \caption[Caption for LOF]{Empirical coverage rates of 95\% confidence intervals for the ATE: Results for Selected Methods
        % --- Begin caption notes block ---
        \par % Start a new line
        \vspace{0.5ex} % Add a small space
        \footnotesize % Use footnote-sized text
        $^1$Abbreviation: Est, Estimated; Pct, Percentile; Par, Parametric. \par % New line
        $^2$The average width of confidence intervals for 10,000 runs of Monte Carlo simulation by each method is provided in the Web Supplementary Material.
        % --- End caption notes block ---
    }
    \label{fig: CProb_Main}
\end{figure}

The empirical coverage rates of the estimated 95\% confidence interval are shown in Figure \ref{fig: CProb_Main}, which has a similar structure of panel to that of Figure \ref{fig: SE.Ratio_Main}. The Y-axis represents the coverage probability of the estimated 95\% confidence interval for each method, with a horizontal line at 95\% indicating the advertised coverage rate. 
%The pattern of clustering defined above for SE estimation is still valid in Figure \ref{fig: CProb_Main}, especially when both the sample proportion of treatment and prevalence of outcome, if subjects are controlled, get neutral to 0.5.
Based on Figure \ref{fig: CProb_Main} and the supplementary material, our results reveal a clear performance hierarchy for CI construction, yielding critical insights in small-sample settings.

The handling of the PS created a fundamental divergence in CI performance. Methods treating the PS as fixed are known to be conservative from asymptotic theory. However, our simulations reveal a noteworthy exception in small-sample settings, where both fixed PS methods (\texttt{Par} \texttt{stdB-Fixed} and \texttt{Pct} \texttt{stdB-Fixed})  exhibited lower coverage rate than their respective re-estimated PS counterparts (\texttt{Par} \texttt{stdB-Est} and \texttt{Pct} \texttt{stdB-Est}). These fixed PS only achieved, and eventually exceeded, the nominal 95\% level as the sample size grew. In stark contrast, accounting for PS estimation led to a dramatic split between asymptotic or bootstrapping approach. The \texttt{Par} \texttt{NS}, despite its theoretical appeal, proved dangerously anti-conservative in practice across all scenarios in small-sample settings. Its failure was exacerbated in scenarios with severe treatment imbalance ($\Pr(Z=1)=10\%$), a condition that challenged all methods but under which \texttt{Par} \texttt{NS} was uniformly the worst performer, with coverage often remaining below 85\% even at $n=1000$. Conversely, bootstrap methods that re-estimated the PS were the most reliable, starting with conservative coverage rates above 95\% in small samples and converging downwards toward the nominal level as the sample size grew, with \texttt{Pct} \texttt{stdB-Est} being particularly effective.

A consistent finding, regardless of whether the PS was treated as fixed or re-estimated, was the relative superiority of the percentile bootstrap over the parametric bootstrap. This performance advantage was particularly pronounced in challenging scenarios of severe treatment imbalance. In these cases, especially at smaller sample sizes, the percentile methods consistently yielded more conservative and reliable coverage rates than its parametric counterpart. This suggests that the percentile method is more robust for constructing confidence intervals when faced with the sparse data conditions common in applied research.

%In summary, for constructing reliable 95\% confidence intervals in small samples, the choice of method is critical. The sandwich estimator with an estimated PS proved to be unreliable, yielding CIs that were consistently too narrow. The most dependable performance was achieved by the bootstrap methods that re-estimated the propensity score. Among these, the percentile bootstrap \texttt{Pct} \texttt{stdB-Est} demonstrated the most robustly conservative behavior across the widest range of challenging scenarios, making it a prudent choice for practical application in studies with limited sample sizes.

\subsection{Comparison of Estimated SE and Confidence Interval based on ATO}

\begin{figure}[h]
    \centering    \includegraphics[width=1\textwidth]{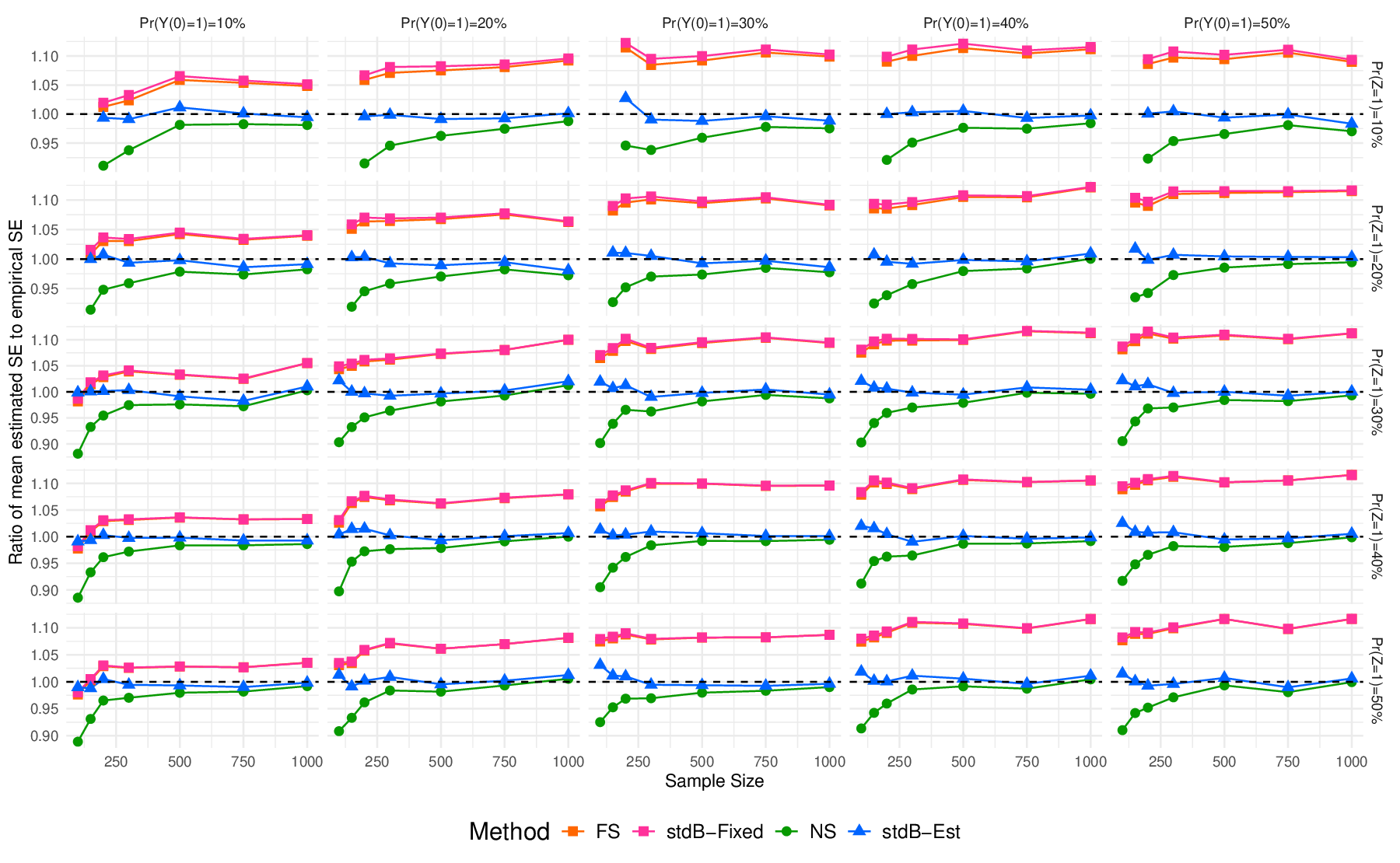}
    \captionsetup{singlelinecheck=off} % Center the caption and allow for footnote
    \caption[Caption for LOF]{Ratio of Mean Estimated SE to Empirical SE for the ATO: Results for Selected Methods
    \par
    \vspace{0.5ex} 
    \footnotesize
    $^1$ The standard errors (SEs) of the estimated SE for 10,000 runs of the Monte Carlo simulation are all less than 0.001 across all methods, demonstrating the precision of our estimated SE results.
    \par
    $^2$ All sandwich variance estimations in this Figure were conducted using the \texttt{PSweight} function from the \texttt{PSweight} R package \cite{zhou2020psweight}.
    }
    \label{fig: SE.Ratio_Main_ATO}
\end{figure}

\begin{figure}[h]
    \centering    \includegraphics[width=1\textwidth]{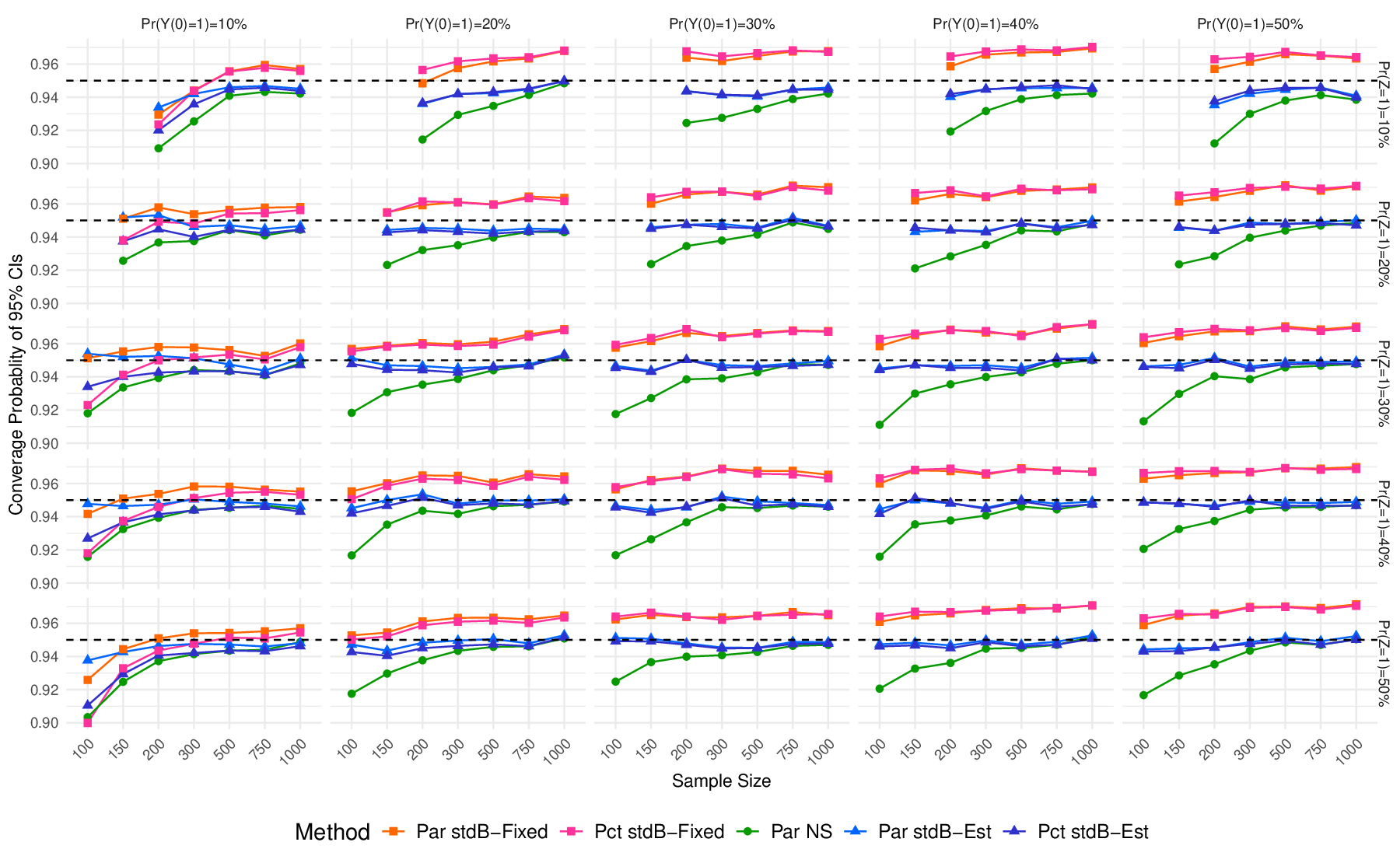}\captionsetup{singlelinecheck=off} %
    \caption[Caption for LOF]{Empirical coverage rates of 95\% confidence intervals for the ATO: Results for Selected Methods
    \par
    \vspace{0.5ex} 
    \footnotesize
    $^1$ Abbreviation: Est, Estimated; Pct, Percentile; Par, Parametric. \par % New line
    $^2$ The average width of confidence intervals for 10,000 runs of Monte Carlo simulation by each method is provided in the Web Supplementary Material.
    }
    \label{fig: CProb_Main_ATO}
\end{figure}

We now evaluate the performance of the estimators using OW. For this analysis, we focus on the same representative methods previously selected for the IPTW results. Specifically, the analysis of the SE ratio, presented in Figure \ref{fig: SE.Ratio_Main_ATO}, includes the same methods as in Section \ref{Sec: RstSE_ATE}. Correspondingly, the analysis of the 95\% CI coverage, shown in Figure \ref{fig: CProb_Main_ATO}, includes the methods from Section \ref{Sec: RstCI_ATE}. These methods were chosen for the main manuscript as the additional methods detailed in the Web Supplementary Material demonstrated either comparable or inferior performance.

As theoretically established\cite{li2019addressing}, OW offers advantages over IPTW by mitigating the influence of extreme PS. Our empirical results confirm this and reveal a critical finding for studies with small sample settings: OW's desirable asymptotic properties manifest at much smaller sample sizes than those required by IPW.  This accelerated convergence means that OW provides more stable and reliable estimates—for both SE ratios and CI coverage—in the small-sample settings that are central to our investigation. The benefit is particularly stark in scenarios with extreme treatment imbalance ($\Pr(Z=1)=10\%$), where OW estimators achieve stable performance with far less data than their struggling IPW counterparts (Figures \ref{fig: SE.Ratio_Main_ATO} vs. \ref{fig: SE.Ratio_Main}).

Specifically under OW, methods treating the PS as fixed (\texttt{FS} and \texttt{stdB-Fixed}) demonstrate this rapid convergence, becoming conservative with SE ratios exceeding one and over nominal 95\% coverage rate even at small sample sizes. By contrast, \texttt{stdB-Est} provides remarkably stable and nearly unbiased SE estimates and stable 95\% coverage rate across all evaluated conditions, achieving its ideal performance almost immediately. Notably, \texttt{NS}, which performed poorly under IPW, shows a dramatic improvement with OW. While it remains the least conservative approach in small samples, its SE ratio rises more swiftly toward one, avoiding the severe underestimation previously observed by IPW.

\subsection{Comparison of Estimated SE and Confidence Interval based on Augmentation with Outcome Model}

\begin{figure}[h]
    \centering    \includegraphics[width=1\textwidth]{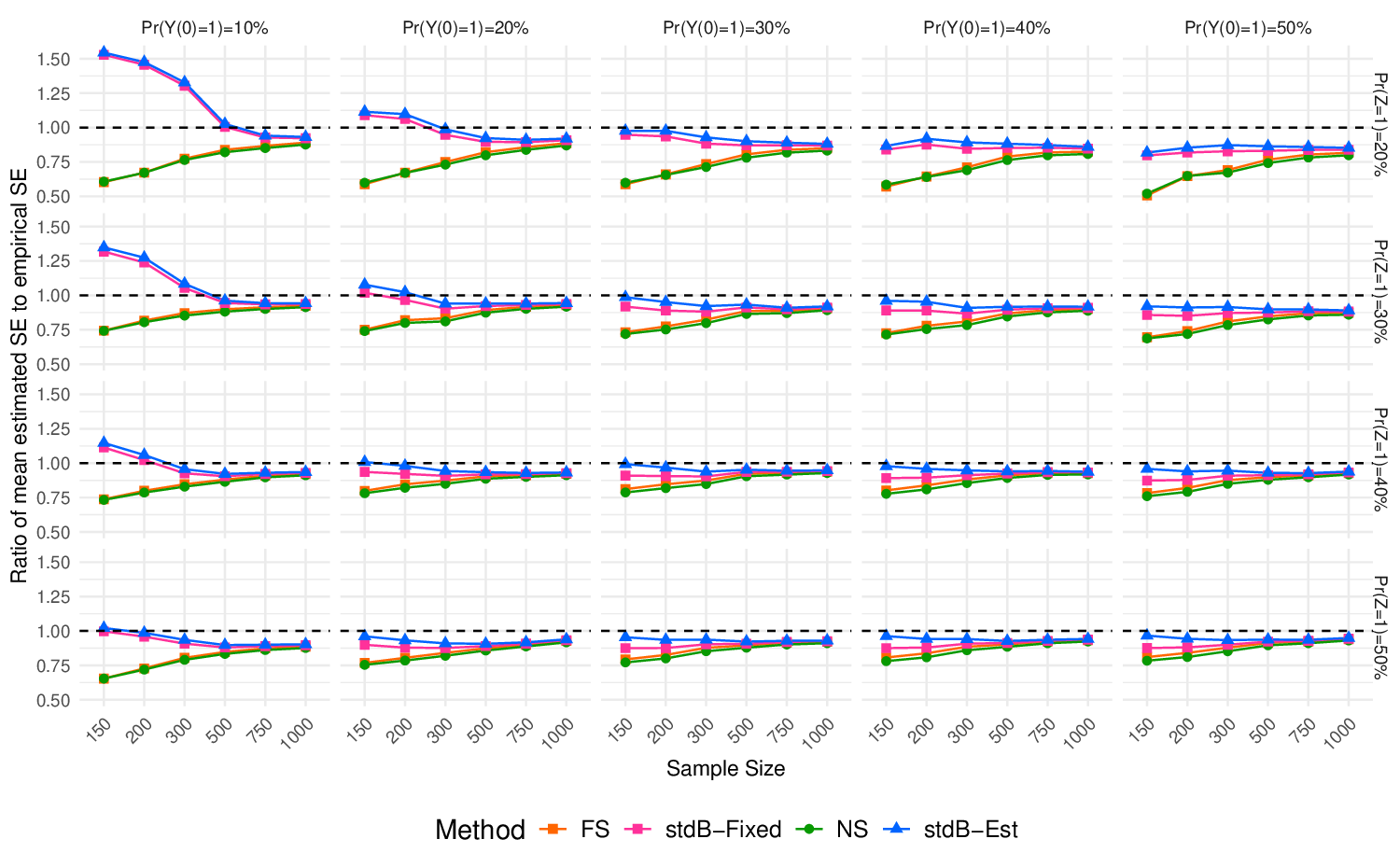}
\captionsetup{singlelinecheck=off} %
    \caption[Caption for LOF]{Ratio of Mean Estimated SE to Empirical SE for the Augmented ATE with \emph{True} Outcome Model: Results for Selected Methods
    \par
    \vspace{0.5ex} 
    \footnotesize
    }
    \label{fig: SE.Ratio_AugT}
\end{figure}

\begin{figure}[h]
    \centering    \includegraphics[width=1\textwidth]{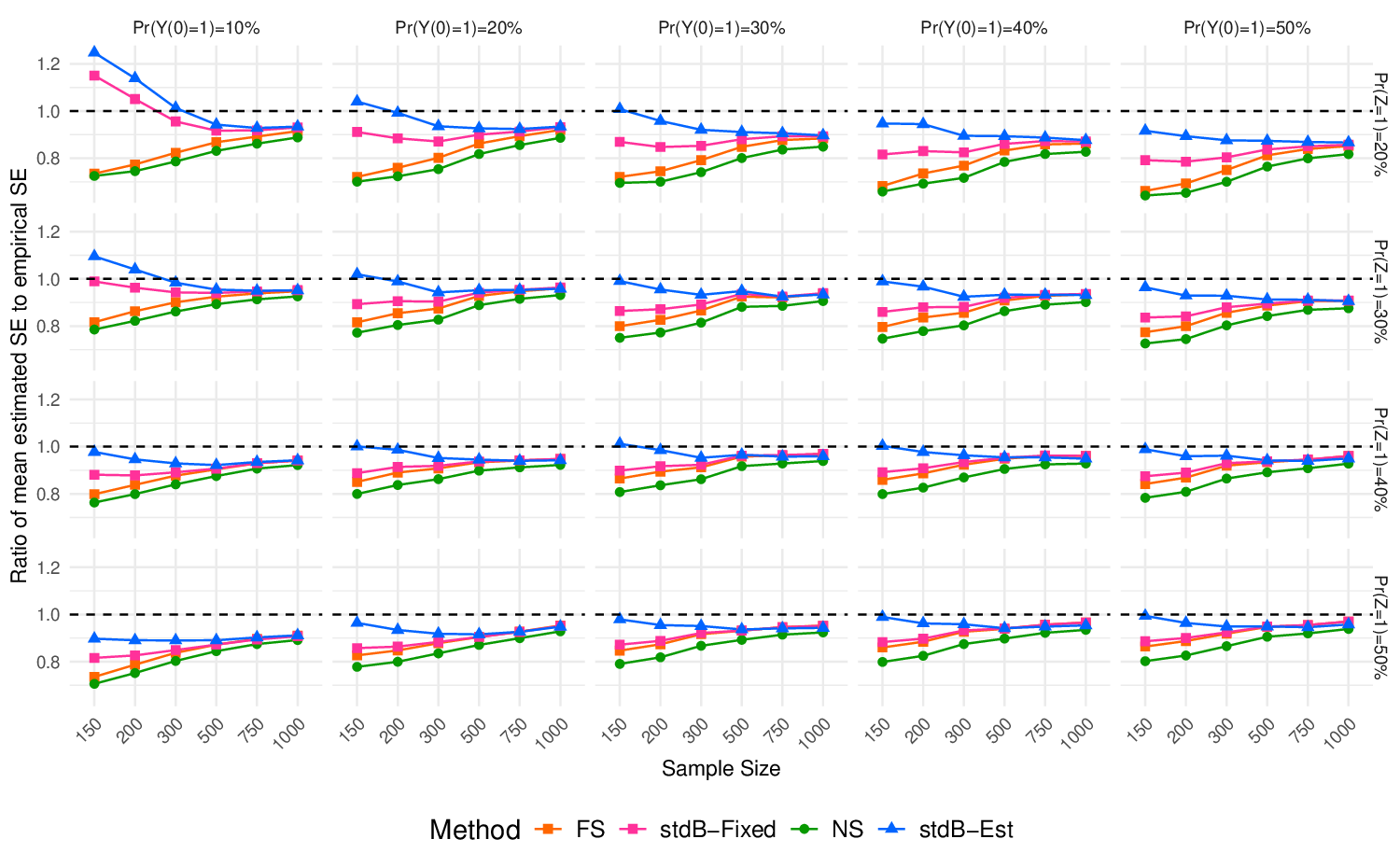}
    \captionsetup{ singlelinecheck=off} %
    \caption[Caption for LOF]{Ratio of Mean Estimated SE to Empirical SE for the Augmented ATE with \emph{Mis-Specified} Outcome Model: Results for Selected Methods
    }
    \label{fig: SE.Ratio_AugF}
\end{figure}

\begin{figure}[h]
    \centering    \includegraphics[width=1\textwidth]{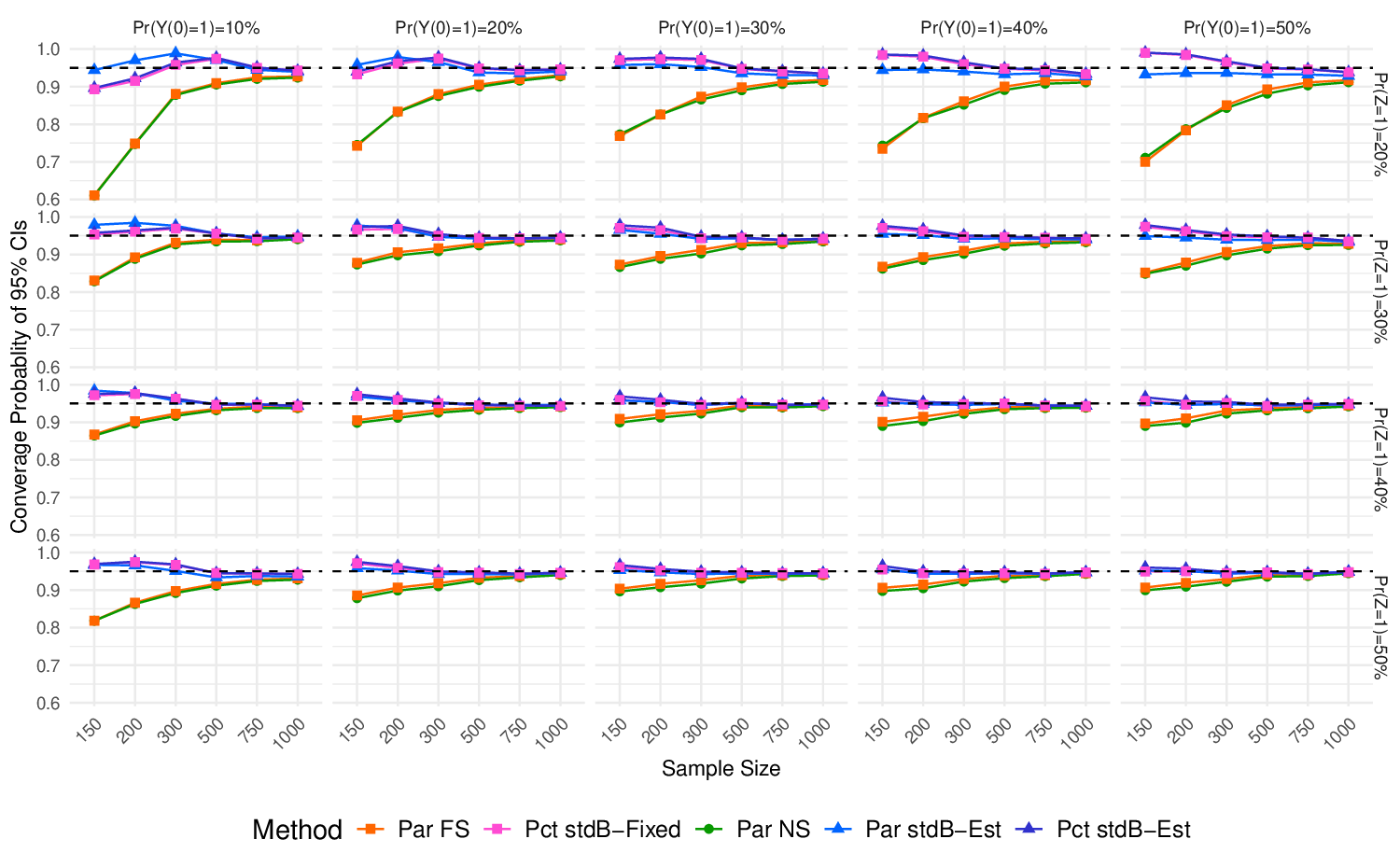}
    \captionsetup{ singlelinecheck=off} %
    \caption[Caption for LOF]{Empirical coverage rates of 95\% confidence intervals among selected Methods which are Augmented by \emph{True} Outcome model}
    \label{fig: CProb_Main_AugT}
\end{figure}

\begin{figure}[h]
    \centering    \includegraphics[width=1\textwidth]{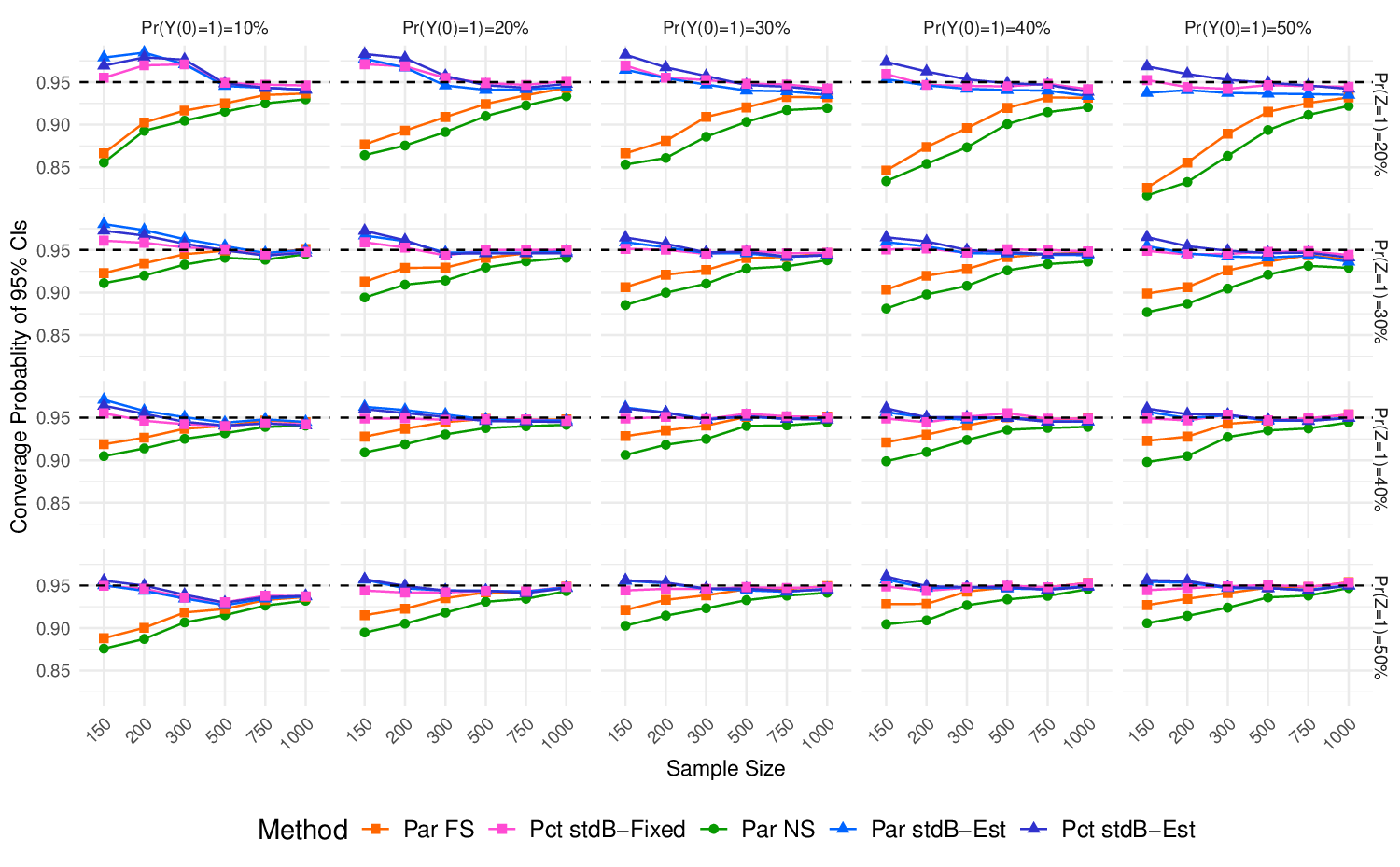}
    \captionsetup{ singlelinecheck=off} %
    \caption[Caption for LOF]{Empirical coverage rates of 95\% confidence intervals among selected Methods which are Augmented by \emph{Mis-specified} Outcome model}
    \label{fig: CProb_Main_AugF}
\end{figure}

In this subsection, we focus on the AIPW estimator. Two outcome models are examined: (1) a correctly-specified outcome model using all covariates \( X_1, X_2, \ldots, X_{10} \); and (2) a mis-specified outcome model using only a subset of five covariates---three continuous variables (\( X_1, X_2, X_3 \)) and two binary variables (\( X_6 \), \( X_{10} \)). Due to quasi-separation issues arising from the low proportion of treatment assignment and outcome events, we start with a sample size of 150 and exclude the scenario with \(\Pr(Z = 1) = 10\%\). We present the ratio of estimated SE to the empirical SE and the corresponding 95\% coverage for selected methods augmented by the correctly-specified outcome model in Figure \ref{fig: SE.Ratio_AugT} and Figure \ref{fig: CProb_Main_AugT}, respectively. Results for methods augmented by mis-specified outcome model are shown in Figure \ref{fig: SE.Ratio_AugF} and Figure \ref{fig: CProb_Main_AugF}. A key aspect of our simulation design must be noted:for calculating the empirical SE (the denominator of our ratio), the point estimate from each Monte Carlo iteration was derived using  non-augmented $\hat{\Delta}_\text{ATE}$ rather than the $\hat{\Delta}_\text{Aug}$. This was done to ensure a consistent benchmark across all methods above. The direct implication of this choice is that the ratio captures both estimator accuracy and efficiency gains. According to asymptotic theory, a correctly specified outcome model yields a more efficient augmented estimator, whose true SE is smaller than our non-augmented benchmark's SE \cite{lunceford2004stratification}. Therefore, when the outcome model is correctly specified (as in Figure \ref{fig: SE.Ratio_AugT}), we expect an accurate SE estimator to produce a ratio below one asymptotically.

As shown in Figure \ref{fig: SE.Ratio_AugT}, the simulation results align perfectly with this theoretical expectation. When the outcome model is correctly specified, the ratio of the mean estimated SE to the empirical SE is uniformly below one for large sample sizes.This observation confirms that the augmented variance estimators are not underestimating. Rather, it validates that they are correctly tracking the smaller, more efficient true variance of the augmented estimator, consistent with asymptotic theory  \cite{lunceford2004stratification}.  

When using the AIPW estimator, all these results indicate that the primary determinant of performance is the choice between bootstrap and asymptotic methods, rendering the handling of the PS. The bootstrap estimators consistently demonstrated more conservative and reliable than their sandwich counterparts across all scenarios, particularly in challenging small sample settings. Their coverage probability always met or exceeded the nominal 95\% level across all scenarios, even in small samples. When comparing within bootstrap approaches, the percentile bootstrap remains the most robustly conservative choice, especially under treatment imbalance and when it is augmented by \textit{mis-specified} outcome model.

In small sample sizes, sandwich estimators tend to substantially underestimate the SE and provide coverage probabilities well below the nominal 95\% level. While this under-estimation diminished as the sample size increased, convergence to nominal performance was notably slow, especially under conditions of imbalanced treatment allocation. A critical and counter-intuitive finding was also observed: augmenting with the true outcome model worsened the performance of the sandwich estimator in small samples, leading to even worse SE under-estimation and lower CI coverage. This occurs because the correctly specified model increases the estimator's efficiency, and the sandwich SE results (Figure \ref{fig: SE.Ratio_AugT} and \ref{fig: SE.Ratio_AugF}) reflect this smaller variance, resulting in dangerously narrow confidence intervals that fail to achieve their nominal coverage rate.

\section{Case Study} \label{Sec: RealAna}
In this section, we present a case study from the LIMIT-JIA Part II study to illustrate the performance differences of the selected methods for SE estimation and 95\% confidence intervals.

\subsection{The LIMIT-JIA study}

The LIMIT-JIA study evaluated whether a 24-dose abatacept regimen, a T-cell co-stimulation inhibitor, in combination with usual care reduced the risk of polyarthritis, uveitis, or the initiation of systemic DMARDs/biologics within 12 months in children with recent-onset limited juvenile idiopathic arthritis (JIA). Limited JIA is defined as arthritis in $\leq$ 4 joints without uveitis, and although it accounts for over half of all JIA cases, it often progresses to polyarthritis (involving $\geq$ 5 joints), uveitis, or necessitates systemic treatment. The primary efficacy outcome is a composite binary endpoint that indicates whether any of the following events occurred within 12 months after enrollment: (1) progression to polyarthritis (i.e., a cumulative total of $\geq$ 5 joints with arthritis since disease onset), (2) onset of uveitis, or (3) initiation of systemic DMARDs and/or biologics, including systemic glucocorticoids.

The LIMIT-JIA study originally began as a randomized controlled trial (Part I), but Part I was terminated due to slow enrollment during the COVID-19 pandemic. Consequently, Part II of the study was modified to a non-randomized format. In Part II, all newly enrolled participants received abatacept. Specifically, 81 eligible participants recruited from CARRA Registry sites were assigned to the abatacept plus usual care treatment arm, while 662 eligible control patients enrolled in the CARRA Registry between its inception in 2015 and May 1, 2023.

A summary of the LIMIT-JIA Part II study comparing the distributions of key covariates in both the unweighted sample and the inverse probability weighted sample is presented in Table~\ref{Tab: LIMIT_Table1}. The weighting procedure is based on PS estimated from a logistic regression model using all seven baseline covariates listed in Table~\ref{Tab: LIMIT_Table1}. Details regarding data processing are provided in the table’s footnote. By applying IPTW, we aim to balance the covariate distributions between the treatment and control groups, thereby mitigating potential confounding. For each subject, the binary primary endpoint was generated using logistic outcome models, resulting in outcome proportions of 15\% in the abatacept + usual care arm and 34.3\% in the usual care arm, respectively. Only the intercept coefficients differ between the outcome models for the control and treatment groups.  Thus, in the LIMIT-JIA Part II study, the proportion of the abatacept + usual care arm is around $\Pr(Z=1) \approx 10\% $, and the proportion of outcomes in the control group is around $\Pr(Y(0)=1) = 34.4\% $. In this way, we can focus on certain cells in the simulation results, Figure \ref{fig: SE.Ratio_Main} and Figure \ref{fig: CProb_Main} as being significant help in determining the performance of the different methods.

The average treatment effect was estimated in LIMIT-JIA along with 95\% confidence intervals based on these four methods: \texttt{Par NS}, \texttt{Par stdB-Est}, \texttt{Par stdB-Fixed}, and \texttt{Pct} \texttt{stdB-Est}. The first three of these are normal-approximation intervals, with the SEs used for their construction also serving as the SE estimators we evaluated. while the \texttt{Pct stdB-Est} method was used only for CI estimation. In addition, we conducted subgroup analyses to explore differences in smaller sample sizes with varying treatment proportions.

\begin{table}[h]
\caption{Baseline Covariate Distributions: Unweighted Versus Inverse Probability Weighted Samples in LIMIT-JIA Part II Study}\label{Tab: LIMIT_Table1}
\centering
\begin{threeparttable}
\begin{tabular}{c|cc|cc}
 & \multicolumn{2}{c|}{Unweighted} & \multicolumn{2}{c}{Weighted} \\ \hline
Counfouder & \begin{tabular}[c]{@{}c@{}}Abatacept + \\      Mean, $P_{50}$, $[P_5,P_{95}]$\end{tabular} & \begin{tabular}[c]{@{}c@{}}Usual Care  \\      Mean, $P_{50}$, $[P_5,P_{95}]$\end{tabular} & \begin{tabular}[c]{@{}c@{}}Abatacept + \\      Mean, $P_{50}$, $[P_5,P_{95}]$\end{tabular} & \begin{tabular}[c]{@{}c@{}}Usual Care  \\      Mean, $P_{50}$, $[P_5,P_{95}]$\end{tabular} \\ \hline
Sex & 32.1\% & 34.4\% & 33.3\% & 34.3\% \\ \hline
Age & 7.6, 6.9, {[}2.5, 15.1{]} & 7.7, 6.7, {[}2.2, 15.2{]} & 7.9, 6.9, {[}2.7, 15.4{]} & 7.7, 6.7, {[}2.2, 15.2{]} \\ \hline
Diagnosis Days & 60.8, 46.0, {[}1.0, 176.0{]} & 53.3, 43.0, {[}1.0, 154.0{]} & 60.8, 46.0, {[}1.0, 166.0{]} & 54.24, 43.0, {[}1.0, 155.0{]} \\ \hline
Uveitis Strata & 59.3\% & 64.4\% & 62.6\% & 64.0\% \\ \hline
Glucocorticoids & 37.0\% & 52.8\% & 56.7\% & 51.2\% \\ \hline
Risk Score & 6.5, 7.0, {[}1.0, 10.0{]} & 4.5, 4.0, {[}0, 10.0{]} & 4.6, 4.0, {[}0, 10.0{]} & 4.7, 4.5, {[}0, 10.0{]} \\ \hline
\# of Active Joints &  &  &  &  \\ \hline
0 & 6.2\% & 34.5\% & 29.8\% & 31.6\% \\
1 & 42.0\% & 43.4\% & 44.1\% & 43.1\% \\
2 & 27.2\% & 15.2\% & 16.6\% & 16.4\% \\
3 & 24.7\% & 6.9\% & 9.4\% & 8.9\%
\end{tabular}
\begin{tablenotes}
    \footnotesize
    \item 1. Glucocorticoid, any intra-articular glucocorticoids used; Risk Score, Juvenile Arthiritis Disease Activity Score;  $P_{50}$, 50th percentile or median; $P_5$, 5th percentile; $P_{95}$, 95th percentile; 
    \item 2. Reference group for binary covariates: Sex, Male as reference group; Uveitis Strata, Lower as reference group; Glucocorticoids, no use of articular glucocorticoids
    \item 3. Multiple imputations are performed using the fully conditional specification method using the pooled Part I and Part II populations for LIMIT-JIA study, implemented via SAS PROC MI, to address the missing covaraites. Analyses for Part II will only use one complete dataset from the 25 imputed complete datasets.
    % \item 4. In the propensity model, treatment status was regressed on the confounder above in Table \ref{Tab: LIMIT_Table1} with an asterisk ($*$). A total of 6 baseline covarites are included in total. Number of active joints at baseline is ordinal categorical variables. The Juvenile arthritis disease activity score was modeled with a restricted cubic spline with six knots, which allows us to flexibly capture non-linear relationships without imposing a strict linearity assumption.
\end{tablenotes}
\end{threeparttable}
\end{table}

\subsection{LIMIT-JIA Study Results}
\begin{figure}[h]
    \centering    \includegraphics[width=\textwidth]{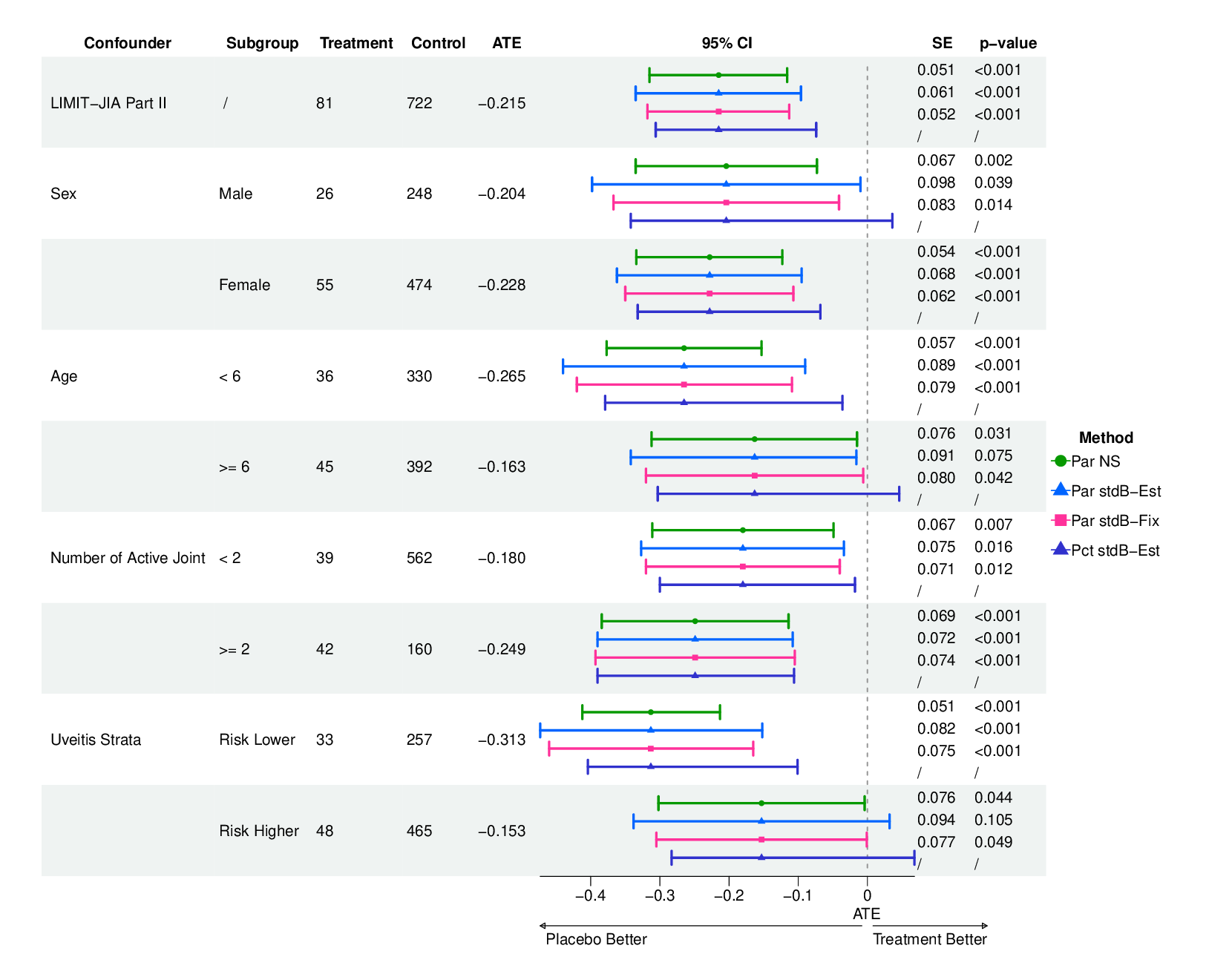}
    \captionsetup{ singlelinecheck=off} %
    \caption[Caption for LOF]{Forest Plot of Subgroup Efficacy in the LIMIT-JIA Part II Study with Selected Methods}
    \label{fig: Eff_Forest}
\end{figure}

The results are summarized in Figure~\ref{fig: Eff_Forest}. We present the forest plot for all patients in the LIMIT-JIA Part II analysis, along with four subgroup analyses based on age ($< 6$ vs.\ $\ge 6$), number of active joints ($< 2$ vs.\ $\ge 2$), and uveitis strata. For each group, the forest plot displays the sample size in each arm, the point estimate of the ATE, its 95\% CI, the estimated SE, and the p-value (if available), using selected methods described above.

From the forest plot, the \texttt{NS} method consistently produces the smallest SE estimates, which in turn yields the narrowest normal-approximation CIs (\texttt{Par NS}) across all subgroup analyses and the overall LIMIT-JIA Part II analysis. Most subgroup analyses yield statistically significant results under all four methods, except the male subgroup, the age group ($\ge 6$), and the higher-risk uveitis strata. In these subgroups, \texttt{Pct} \texttt{stdB-Est} consistently provides the most conservative estimates. 

Considering the simulation performance shown in Figure~\ref{fig: SE.Ratio_Main} and Figure~\ref{fig: CProb_Main}, results based on \texttt{NS} and \texttt{Par NS} are the least reliable. For example, in the subgroup analysis of the higher-risk uveitis strata, we recommend relying on \texttt{stdB-Est} and \texttt{Pct} \texttt{stdB–Est} for drawing more conservative and trustworthy conclusions.

\section{Discussion} \label{Sec: Discussion}

%In summary, our findings highlight a clear performance hierarchy among the four selected methods, particularly in small-sample settings (e.g., $n \le 200$) where all estimators exhibited some degree of bias. Across the scenarios we evaluated, \texttt{stdB-Est} was consistently the most conservative method. In stark contrast, \texttt{NS} was uniformly the least conservative, frequently underestimating the true variability regardless of sample size. Notably, the two methods that treated the PS as fixed, which are known from asymptotic theory to be conservative, were in fact less conservative than \texttt{stdB-Est} in these small-sample scenarios. This suggests that for studies with limited sample sizes, a bootstrap procedure that fully accounts for all sources of estimation uncertainty is the most prudent approach for ensuring conservative statistical inference.

In this study, we systematically evaluated the performance of sandwich and bootstrap variance estimators and the reliability of their corresponding CIs for propensity score-weighted analyses, with a specific focus on the small-sample settings. We conducted an extensive Monte Carlo simulation to compare methods based on how the PS was handled (fixed vs re-estimated), the choice of weighting scheme (IPTW vs OW), the use of outcome model augmentation (AIPW) and the method of CI construction (e.g., parametric vs percentile bootstrap). By examing performance across various sample sizes, treatment allocation ratio, and outcome prevalences, we aimed to provide clear, evidence-based guidance for researchers working with limited data.  

Our findings challenge several common assumptions that underpin PS analysis, particularly when applied to small samples. First, we address the common assumption, documented in the literature \cite{reifeis2022variance, mao2019propensity, kostouraki2024variance}, that treating the PS as fixed tends to produce conservative (i.e., overestimated) variance estimates. Our simulations demonstrate that this asymptotic property does not reliably hold in small-sample settings, where this method is frequently anti-conservative. This distinction between theory and small-sample practice represents a critical risk for applied researchers. Second, our results question the practical utility of theoretically-correct asymptotic estimators, such as the \texttt{NS} method. Despite properly accounting for PS estimation, its performance under IPTW was dangerously anti-conservative, underscoring that its reliability is highly dependent on the dual conditions of both sufficient sample size and balanced treatment allocation. 

Crucially, our study not only identifies these practical challenges but also provides clear, evidence-based solutions. For standard IPTW, the primary recommendation for analyses with small samples or imbalanced data is to abandon unreliable asymptotic methods and instead use the bootstrap. The \texttt{stdB-Est} consistently provided stable SE estimates, while the \texttt{Pct} \texttt{stdB-Est} proved superior for constructing confidence intervals with robustly conservative coverage.

The use of OW offers a powerful alternative, although this changes the estimand which focuses on the subset of subjects with the most clinical equipoise. The key finding for OW is its accelerated convergence to asymptotic behavior. By naturally bounding the weights, OW mitigates the high variance that plagues IPTW, allowing all methods to perform reliably at much smaller sample sizes. This demonstrates that the choice of weighting scheme is not merely a preference but a critical determinant of estimator stability.

Furthermore, our investigation into AIPW—often considered a more advanced, robust estimator—revealed a critical cautionary tale for small-sample inference. Attempting to improve precision by augmenting with a correctly specified outcome model paradoxically worsened the performance of the sandwich estimator, leading to even more severe SE underestimation and CI under-coverage. In contrast,  \texttt{stdB-Est} remained remarkably robust. \texttt{Pct} \texttt{stdB-Est}, in particular, maintained reliable coverage even when the outcome model was mis-specified, underscoring its superior resilience and making it the most dependable choice when employing augmentation in practice.

Based on our findings, we offer clear guidance for PS analysis in small samples. The bootstrap with PS re-estimation (\texttt{stdB-Est}) should be the default choice for both SE estimation and CI construction, as it consistently provides reliable and appropriately conservative results where asymptotic methods fail. For CIs specifically, the percentile bootstrap (\texttt{Pct} \texttt{stdB-Est}) is recommended for its superior robustness, particularly when using AIPW or in cases of treatment imbalance. When extreme PS are a concern, analysts may consider OW over IPTW to ensure estimator stability, while acknowledging the change in estimand to the ATO. We issue a strong caution against the use of sandwich estimators in small-sample contexts; they are systematically anti-conservative, and with AIPW, can paradoxically produce misleadingly narrow confidence intervals even when the outcome model is correct.

A critical next step is to extend out small-sample investigation to other data types, particularly survival endpoints. While foundational work exists - such as Austin (2016)\cite{austin2016variance}, who championed the bootstrap for weighted Cox models, and Hajage et al. (2018)\cite{hajage2018closed}, who developed a theoretically-correct closed-form variance estimator - these studies have crucial limitations. Austin's sandwich estimator for comparison assumed fixed PS, while method of Hajage et al. was validated only in large-sample settings ($\geq 2,000$). Consequently, the performance of these methods in the small-sample scenarios central to our paper remains a significant and unaddressed knowledge gap. Furthermore, given our findings on the  superiority of the percentile bootstrap for binary outcomes, a comprehensive evaluation of alternative CI construction methods beyond the standard normal-based interval is a vital avenue for future research to refine best practices for causal inference with limited time-to-event data.

\section{Acknowledgement}
Laine E Thomas, Ph.D., holds an Innovation in Regulatory Science Award from the Burroughs Wellcome Fund. The LIMIT-JIA example was supported by the PCORI funded project (ID 8177) Improving Outcomes in Limited Juvenile Idiopathic Arthritis ClinicalTrials.gov ID: NCT03841357.
The contents of this article are solely the responsibility of the authors and do not necessarily represent the view of PCORI.  We appreciate the clinical input and motivating questions from LIMIT-JIA PI Laura Schaunberg.  

\section{Data Availability}
R programming used to develop the study can be found on
GitHub: \url{https://github.com/BaoshanZZ/VarEst_PS}.

\bibliography{ref}

\newpage

\section{Appendix}

\subsection{Purely Empirical Sandwich Estimator for ATE}

In this section, we develop the purely empirical sandwich estimator (($\hat{\Sigma}_\text{PES}$)) for the Average Treatment Effect (ATE) derived from a system of estimating equations. We also delineate the key differences between this robust estimator and a commonly referenced-based model sandwich estimator ($\hat{\Sigma}_\text{MS}$) attributed to \cite{lunceford2004stratification}, which exploits the PS model assumption for simplification. 

Consider the stacked estimating function $\psi_i(\theta)$ for subject $i$, designed to jointly estimate the propensity score (PS) parameters ($\beta$), the potential outcome means ($\mu_1, \mu_0$), and the ATE ($\Delta_\text{ATE}$):
\[
    \psi_i(\theta) = \begin{bmatrix}
    \psi_i^\beta(\beta) \\
    \psi_i^{\mu_1}(\beta, \mu_1) \\
    \psi_i^{\mu_0}(\beta, \mu_0) \\
    \psi_i^\Delta(\mu_1, \mu_0, \Delta)
    \end{bmatrix}
    =
    \begin{bmatrix}
    (Z_i - e(\mathbf{X}_i, \beta)) \mathbf{X}_i \\
    \frac{Z_i (Y_i - \mu_1)}{e(\mathbf{X}_i, \beta)}\\
    \frac{(1-Z_i) (Y_i - \mu_0)}{1 - e(\mathbf{X}_i, \beta)} \\
    \mu_1-\mu_0-\Delta
    \end{bmatrix}
    :=
    \begin{bmatrix}
    (Z_i - e_i) \mathbf{X}_i \\
    \frac{Z_i (Y_i - \mu_1)}{e_i}\\
    \frac{(1-Z_i) (Y_i - \mu_0)}{1 - e_i} \\
    \mu_1-\mu_0-\Delta
    \end{bmatrix}
\]
The full parameter vector is $\theta = (\beta^\top, \mu_1, \mu_0, \Delta)^\top \in \mathbb{R}^{p+3}$. The estimator $\hat{\theta}$ is obtained by solving $\sum_{i=1}^n \psi_i(\hat{\theta}) = \mathbf{0}$.

To construct the sandwich variance estimator, we require the "bread" matrix $\mathbf{A}(\theta) = -E[\dot{\psi}_i(\theta)]$ and the "meat" matrix $\mathbf{B}(\theta) = E[\psi_i(\theta)\psi_i(\theta)^\top]$, where $\dot{\psi}_i(\theta) = \partial \psi_i(\theta) / \partial \theta^\top$. Assuming a logistic PS model, the negative derivative matrix for a single observation is , $e_i = \text{expit}(\mathbf{X}_i^\top \beta)$, we have $\partial e_i / \partial \beta = e_i(1-e_i)\mathbf{X}_i$. Substituting this gives:
\[
-\dot{\psi_i}=
\left[\begin{array}{c|cc|c}
 e_i(1-e_i)\mathbf{X}_i\mathbf{X}_i^\top & 0& 0&0 \\ \hline

 \frac{Z_i(Y_i-\mu_1)(1-e_i)\mathbf{X}_i^\top}{e_i} & \frac{Z_i}{e_i} & 0& 0\\

-\frac{(1-Z_i)(Y_i-\mu_0)e_i\mathbf{X}_i^\top}{1-e_i} &0 & \frac{1-Z_i}{1-e_i} &0 \\ \hline

0& -1 & 1 & 1

 \end{array}\right].
\]
The outer product matrix $\psi_i(\theta)\psi_i(\theta)^\top$, when evaluated at the solution $\hat{\theta}$ where $\psi_i^\Delta(\hat{\theta}) = \hat{\mu}_1 - \hat{\mu}_0 - \hat{\Delta} = 0$, becomes:
\[
    \psi_i(\hat{\theta})\psi_i(\hat{\theta})^\top = 
 \left[\begin{array}{c|cc|c}

 (Z_i-e_i)^2\mathbf{X}_i\mathbf{X}_i^\top &\frac{Z_i(Y_i-\mu_1)(1-e_i)\mathbf{X}_i}{e_i} & -\frac{(1-Z_i)(Y_i-\mu_0)e_i\mathbf{X}_i}{1-e_i} & 0 \\ \hline

 \frac{Z_i(Y_i-\mu_1)(1-e_i)\mathbf{X}_i^\top}{e_i} & \frac{Z_i(Y_i-\mu_1)^2}{e_i^2} & 0 & 0\\

 -\frac{(1-Z_i)(Y_i-\mu_0)e_i\mathbf{X}_i^\top}{1-e_i} & 0 & \frac{(1-Z_i)(Y_i-\mu_0)^2}{(1-e_i)^2} & 0 \\ \hline
   0 & -1 & 0&    
 \end{array}\right]
\]
Note that the off-diagonal terms in the middle block $(2,3)$ and $(3,2)$ are $\frac{Z_i(1-Z_i)(Y_i-\hat{\mu}_1)(Y_i-\hat{\mu}_0)}{\hat{e}_i(1-\hat{e}_i)}$, which is always zero since $Z_i(1-Z_i)=0$.

Following the notation from  \cite{stefanski2002calculus}, the purely empirical estimators (PSE) for $\mathbf{A}(\theta_0)$ and $\mathbf{B}(\theta_0)$ are constructed using empirical averages evaluated at $\hat{\theta}$:
\[
    \mathbf{\bar{A}}_n = \frac{1}{n}\sum_{i=1}^n[-\dot{\psi_i}(\hat{\theta})] \quad \text{and} \quad \mathbf{\bar{B}}_n = \frac{1}{n}\sum_{i=1}^n[\psi_i(\hat{\theta})\psi_i(\hat{\theta})^\top]
\]
Using block matrix notation corresponding to the partition $(\beta^\top \mid \mu_1, \mu_0 \mid \Delta)^\top$:
\
where
\begin{itemize}
    \item $\bar{\mathbf{A}}_{11} = \frac{1}{n}\sum_{i=1}^n \hat{e}_i(1-\hat{e}_i)\mathbf{X}_i\mathbf{X}_i^\top$
    \item $\bar{\mathbf{H}} = \frac{1}{n}\sum_{i=1}^n \begin{bmatrix} \frac{Z_i(Y_i-\hat{\mu}_1)(1-\hat{e}_i)\mathbf{X}_i^\top}{\hat{e}_i} \\ -\frac{(1-Z_i)(Y_i-\hat{\mu}_0)\hat{e}_i\mathbf{X}_i^\top}{1-\hat{e}_i} \end{bmatrix}$
    \item $\bar{\mathbf{A}}_{22} = \frac{1}{n}\sum_{i=1}^n \begin{bmatrix} \frac{Z_i}{\hat{e}_i} & 0 \\ 0 & \frac{1-Z_i}{1-\hat{e}_i} \end{bmatrix}$
    \item $\mathbf{C} = [1, -1]$
    \item $\bar{\mathbf{B}}_{11} = \frac{1}{n}\sum_{i=1}^n (Z_i-\hat{e}_i)^2\mathbf{X}_i\mathbf{X}_i^\top$
    \item $\bar{\mathbf{B}}_{12} = \frac{1}{n}\sum_{i=1}^n (Z_i-\hat{e}_i)\mathbf{X}_i \left[ \frac{Z_i(Y_i-\hat{\mu}_1)}{\hat{e}_i} \quad \frac{(1-Z_i)(Y_i-\hat{\mu}_0)}{1-\hat{e}_i} \right]$
    \item $\bar{\mathbf{B}}_{21} = \bar{\mathbf{B}}_{12}^\top$
    \item $\bar{\mathbf{B}}_{22} = \frac{1}{n}\sum_{i=1}^n \begin{bmatrix} \frac{Z_i(Y_i-\hat{\mu}_1)^2}{\hat{e}_i^2} & 0 \\ 0 & \frac{(1-Z_i)(Y_i-\hat{\mu}_0)^2}{(1-\hat{e}_i)^2} \end{bmatrix}$
\end{itemize}

The inverse of the block lower triangular matrix $\mathbf{\bar{A}}_n$ is also block lower triangular. The third block row, corresponding to the parameter $\Delta_\text{ATE}$, is given by:
\[
    (\mathbf{\bar{A}}_n^{-1})_{3.} = \left[
    \begin{array}{c|c|c}
    -\mathbf{C}(\bar{\mathbf{A}}_{22})^{-1}\bar{\mathbf{H}}(\bar{\mathbf{A}}_{11})^{-1} & \mathbf{C}(\bar{\mathbf{A}}_{22})^{-1}& 1
    \end{array}\right]
\]
The purely empirical sandwich variance estimator for $\hat{\Delta}$ is the $(3,3)$ element of the matrix $n^{-1} \mathbf{\bar{A}}_n^{-1} \mathbf{\bar{B}}_n (\mathbf{\bar{A}}_n^{-1})^\top$. This can be expressed as the empirical variance of the estimated influence function for $\hat{\Delta}$:
\begin{equation}
    \mathbf{V}_n( \hat{\Delta}_{\text{ATE}}) = (\mathbf{\bar{A}}_n^{-1})_{3.} (\bar{\mathbf{B}}) (\mathbf{\bar{A}}_n^{-1})_{3.}^\top = \frac{1}{n}\sum_{i=1}^n\left\{\mathbf{C}(\bar{\mathbf{A}}_{22,i})^{-1}(\mathbf{\psi}_i^\mu - \bar{\mathbf{H}}(\mathbf{\bar{A}}_{11})^{-1}\psi_i^\beta\right\},
\end{equation}
where $\psi_i^\mu = [\psi_i^{\mu_1}(\hat{\theta}), \psi_i^{\mu_0}(\hat{\theta})]^\top$. This estimator uses only empirical averages and does not rely on further model assumptions beyond the structure of the estimating equations themselves. Its key advantage is robustness: it provides asymptotically valid variance estimates even if the PS model $e(\mathbf{X}, \beta)$ is mis-specified \cite{stefanski2002calculus}.

We now contrast this PSE with the sandwich estimator proposed by \cite{lunceford2004stratification}. Their approach, particularly as described in relation to Formula 19 of their paper, can be interpreted as simplifying the $\mathbf{\bar{A}}_n$ matrix by exploiting the assumption that the PS model is correctly specified. Specifically, under the assumption $\E[Z_i | \mathbf{X}_i] = e_i$, the expectation of the diagonal elements of $\bar{\mathbf{A}}_{22}$ is $\E[\text{diag}(Z_i/e_i, (1-Z_i)/(1-e_i))] = \mathbf{I}_2$, the $2 \times 2$ identity matrix. The \cite{lunceford2004stratification} estimator, in this simplified form, replaces the empirical block $\bar{\mathbf{A}}_{22}$ with its theoretical expectation $\mathbf{I}_2$ in the calculation of $\mathbf{\bar{A}}_n^{-1}$. 

This substitution leads to a simplified expression for the third row of the inverse "bread" matrix:
\[
    (\mathbf{A}_{n, ~ \text{L\&D}}^{-1})_{3.} = \left[
    \begin{array}{c|c|c}
    -\mathbf{C}\bar{\mathbf{H}}(\bar{\mathbf{A}}_{11})^{-1} & \mathbf{C} & 1
    \end{array}\right]
\]
Consequently, the resulting variance estimator for $\hat{\Delta}$ differs from the PSE. 
% While potentially more efficient in finite samples \emph{if} the PS model is truly correct, this simplified estimator loses the robustness property. If the PS model is mis-specified, replacing $\bar{\mathbf{A}}_{22}$ with $\mathbf{I}_2$ is incorrect, leading to an inconsistent variance estimator and potentially invalid statistical inference (e.g., confidence intervals with incorrect coverage). 
 
The estimator discussed by \cite{lunceford2004stratification}, while related, represents a version that achieves simplification by relying on the correctness of the PS model, trading robustness for potential efficiency gains under that specific assumption. It is important to note that the L\&D variance estimator is just one example among a range of possibilities where specific model assumptions, such as the correct specification of the PS, are exploited to simplify the variance estimation procedure \cite{stefanski2002calculus}. The choice between the robust PSE and such model-dependent estimators depends on the confidence in the underlying model specification, with the PSE being the more conservative option when model mis-specification is a concern.

In summary, the purely empirical sandwich estimator (PSE) derived here provides a robust method for estimating the variance of the IPW-based ATE estimator, accounting for the uncertainty in PS estimation without requiring the assumption that the PS model is correctly specified. The estimator discussed by Lunceford and Davidian (2004), while related, represents a version that achieves simplification by relying on the correctness of the PS model, trading robustness for potential efficiency gains under that specific assumption \cite{lunceford2004stratification}. The choice between them depends on the confidence in the PS model specification, with the PSE being the more conservative and robust option when model misspecification is a concern.

\subsection{Full Simulation Results}

A central finding from the full Monte Carlo simulations is that all evaluated variance estimation methods consistently partition into three distinct performance clusters. This robust clustering, defined by the  is defined by the methods' core theoretical framework—which models the interaction between variance estimation methods and PS estimation— and serves as the primary organizing principle for all results in this supplemantary material. We observed this consistent clustering phenomenon not only in the ATE SE estimation results (Figure \ref{fig: SE.Ratio_Full}) but also in ATE confidence interval (CI) coverage (Figure \ref{fig: CProb_Full}) and all corresponding results for the ATO (Figures \ref{fig: SE.Ratio_Full_ATO} -- \ref{fig: Width_Full_ATO}).

The three identified clusters are defined as follows:
\begin{enumerate}
    \item \textbf{Cluster 1: Fixed PS (\texttt{Fixed})}: Includes all methods, both asymptotic and bootstrap, that treat the PS as a fixed, known quantity, thereby ignoring estimation uncertainty. (Marked as $\square$ with \textcolor{red}{red} in figures).
    \item \textbf{Cluster 2: Estimated PS - Sandwich (\texttt{Sandwich-Est})}: Includes all asymptotic (sandwich) variance methods that analytically account for PS model estimation. (Marked as $\bigcirc$ with \textcolor{ForestGreen}{green} in figures).
    \item \textbf{Cluster 3: Estimated PS - Bootstrap (\texttt{Bootstrap-Est})}: Includes all bootstrap methods where the PS model is re-estimated within each bootstrap resample to capture estimation uncertainty. (Marked as $\triangle$ with \textcolor{blue}{blue} in figures).
\end{enumerate}

This consistent, three-group structure strongly guided our selection of representative methods for the main manuscript. To clearly illustrate the performance differences between these distinct approaches, we selected popular and high-performing methods from each cluster: \texttt{FS} and \texttt{stdB-Fixed} from the \textbf{Fixed} cluster, \texttt{NS} from the \textbf{Sandwich-Est} cluster, and \texttt{stdB-Est} from the \textbf{Bootstrap-Est} cluster. The following sections detail the performance trends observed within and between these clusters, beginning with the ATE.

\subsubsection{Ratio of mean estimated SE to empirical SE for ATE}

We begin by examining the ratio of the mean estimated SE to the empirical SE (i.e., the standard deviation of the 10,000 point estimates). This metric, which should ideally be 1, provides the clearest initial demonstration of the three performance clusters.

\begin{figure}[h]
    \centering    
    \includegraphics[width = \textwidth]{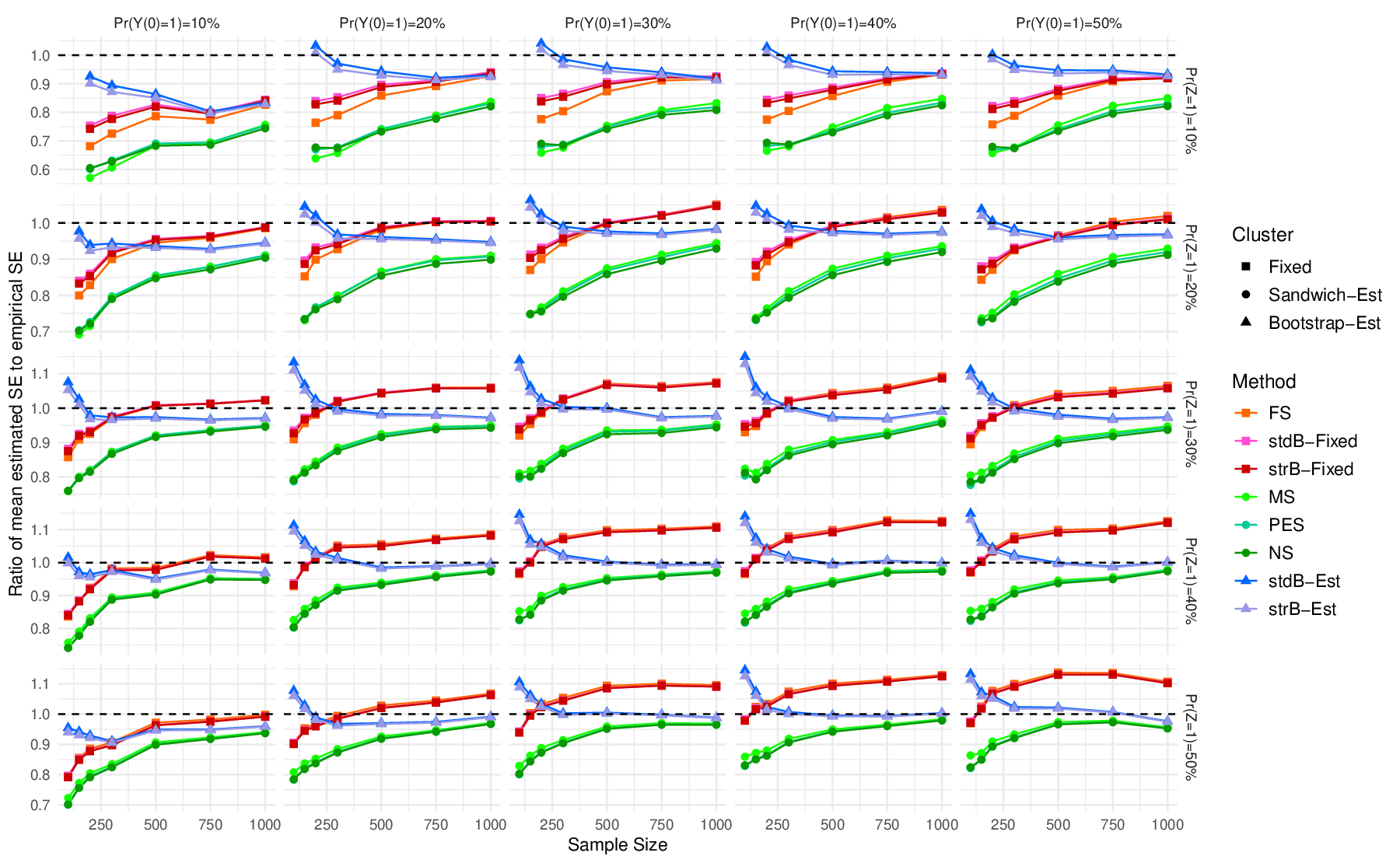} \captionsetup{justification=centering, singlelinecheck=off} % Center the caption and allow for footnote
    \caption[Caption for LOF]{Ratio of mean estimated SE\footnotemark to empirical SE for the ATE: Results for all Methods, illustrating the three performance clusters.}
    \label{fig: SE.Ratio_Full}
\end{figure}
\footnotetext{The SEs of the mean estimated SE for 10,000 runs of the Monte Carlo simulation are all less than 0.001 across all methods, demonstrating the precision of our estimated SE results.}

As shown in Figure \ref{fig: SE.Ratio_Full}, methods within each cluster exhibit highly similar performance trends as sample size increases. Our analysis of these clusters reveals several key findings:

\paragraph{Trend of SE Estimation within Each Cluster:}
\begin{itemize}
    \item \textbf{Cluster \texttt{Fixed}}: This cluster generally shows an increasing SE ratio as the sample size increases. In most scenarios, the ratio starts below one (anti-conservative) in small samples and exceeds one (overly conservative) as sample size grows. This confirms the known theoretical property that treating the PS as fixed leads to an overestimation of the true variance in large samples \cite{lunceford2004stratification}. Notably, in small-sample settings, \texttt{stdB-Fixed} and \texttt{strB-Fixed} are consistently a more conservative variance estimator than \texttt{FS}, although the performance difference between these two bootstrap approaches was negligible.s all sample settings. 
    \item \textbf{Cluster \texttt{Sandwich-Est}}: This cluster also exhibits an increasing trend, starting below one (anti-conservative) and approaching the target ratio of one as sample size grows. However, it consistently yields the least conservative (smallest) SE estimates among all methods, particularly in small samples where its SE ratio is the lowest. The performance of individual methods within this cluster was nearly identical, making them practically indistinguishable in our settings.
    \item \textbf{Cluster \texttt{Bootstrap-Est}}: In contrast, methods in this cluster start with the most conservative SE ratios in small samples (often $>$ 1) and correctly converge toward the target ratio of one in large samples. Within this cluster, \texttt{stdB-Est} consistently shows slightly more conservative SEs than \texttt{stfB-Est}.
\end{itemize}

\paragraph{Comparison of SE Estimation between Clusters:}
\begin{itemize}
    \item \textbf{\texttt{Fixed} vs. \texttt{Sandwich-Est}}: Across all sample sizes, the \textbf{\texttt{Fixed}} cluster yields uniformly larger SE estimates than the \textbf{\texttt{Sandwich-Est}} cluster. This empirically confirms that accounting for PS estimation via the sandwich estimator (\texttt{Sandwich-Est}) correctly results in a smaller, more efficient variance estimate.
    \item \textbf{\texttt{Sandwich-Est} vs. \texttt{Bootstrap-Est}}: The \textbf{\texttt{Bootstrap-Est}} methods produce larger (more conservative) SE estimates than the \textbf{\texttt{Sandwich-Est}} methods, especially in small samples. This highlights the conservative tendency of the bootstrap in small $n$ settings. This difference diminishes as the sample size increases, with both clusters converging toward the target ratio of one.
    \item \textbf{\texttt{Bootstrap-Est} vs. \texttt{Fixed}}: The relationship between these two clusters inverts with sample size. In small samples, the \textbf{\texttt{Bootstrap-Est}} cluster is more conservative. As the sample size increases, the \textbf{\texttt{Fixed}} cluster becomes more conservative, overshooting the target ratio of one while the \textbf{\texttt{Bootstrap-Est}} cluster converges correctly to it.
\end{itemize}

\begin{figure}
    \centering
    \includegraphics[scale = 0.5]{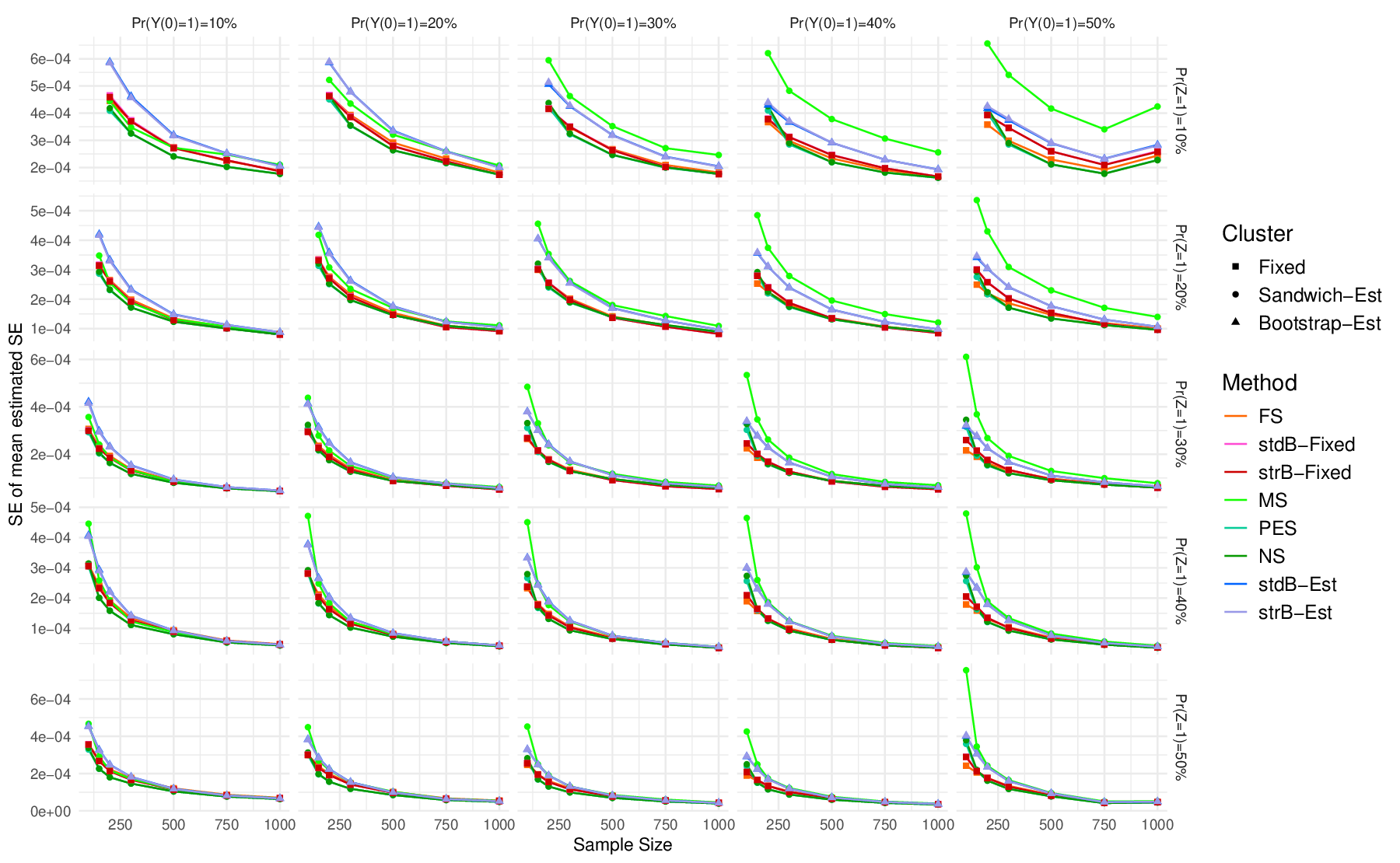}
    \caption{Standard error of all estimated SEs among Monte Carlo Simulation for ATE: Results for all methods}
    \label{fig: SE.SE_Full}
\end{figure}

\subsubsection{Empirical Coverage Probability for ATE}
\begin{figure}[h]
    \centering    \includegraphics[scale = 0.55]{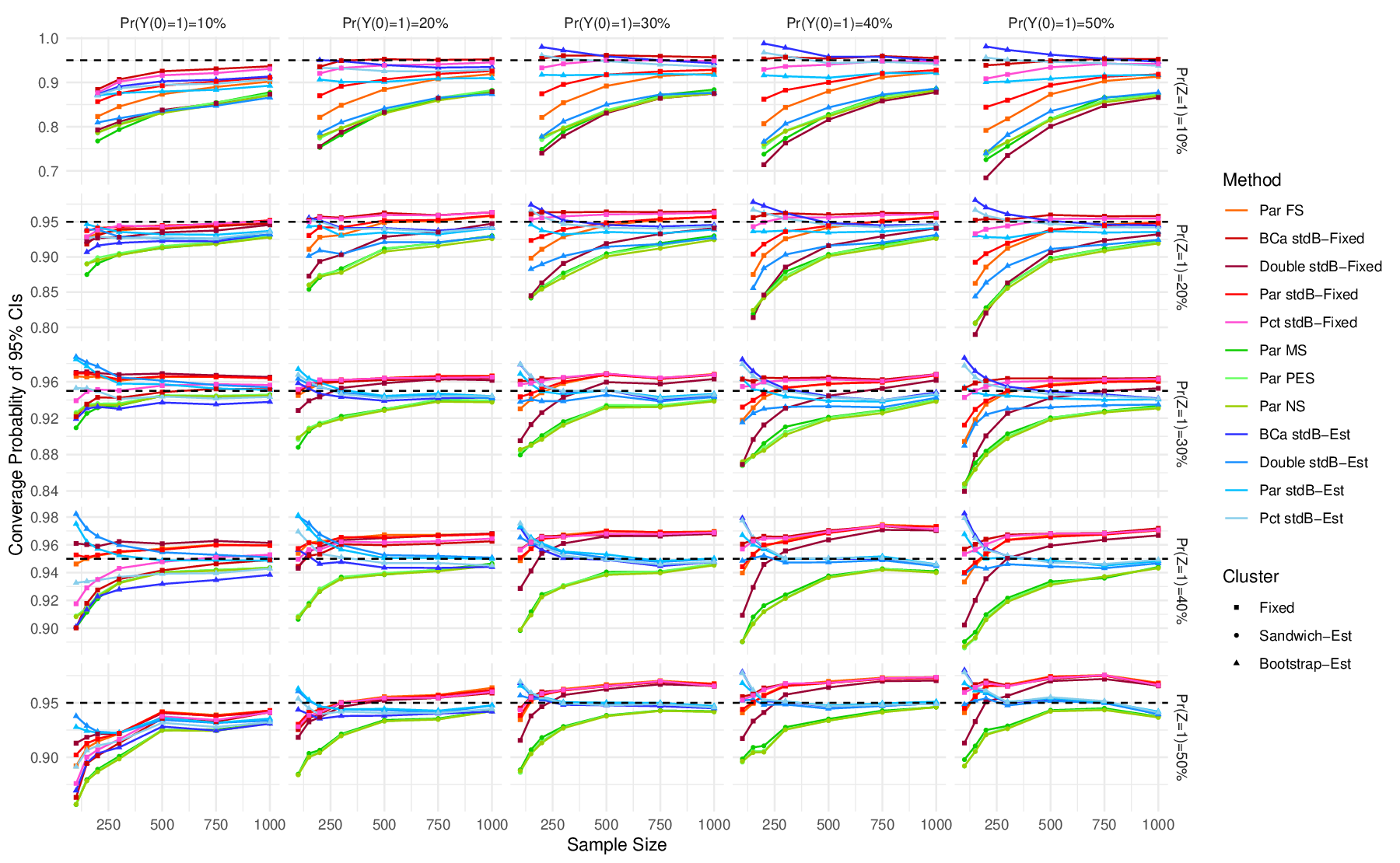}
    \caption{Empirical coverage rates of 95\% confidence intervals for the ATE: Results for all Methods}
    \label{fig: CProb_Full}
\end{figure}

\begin{figure}[h]
    \centering    \includegraphics[scale = 0.55]{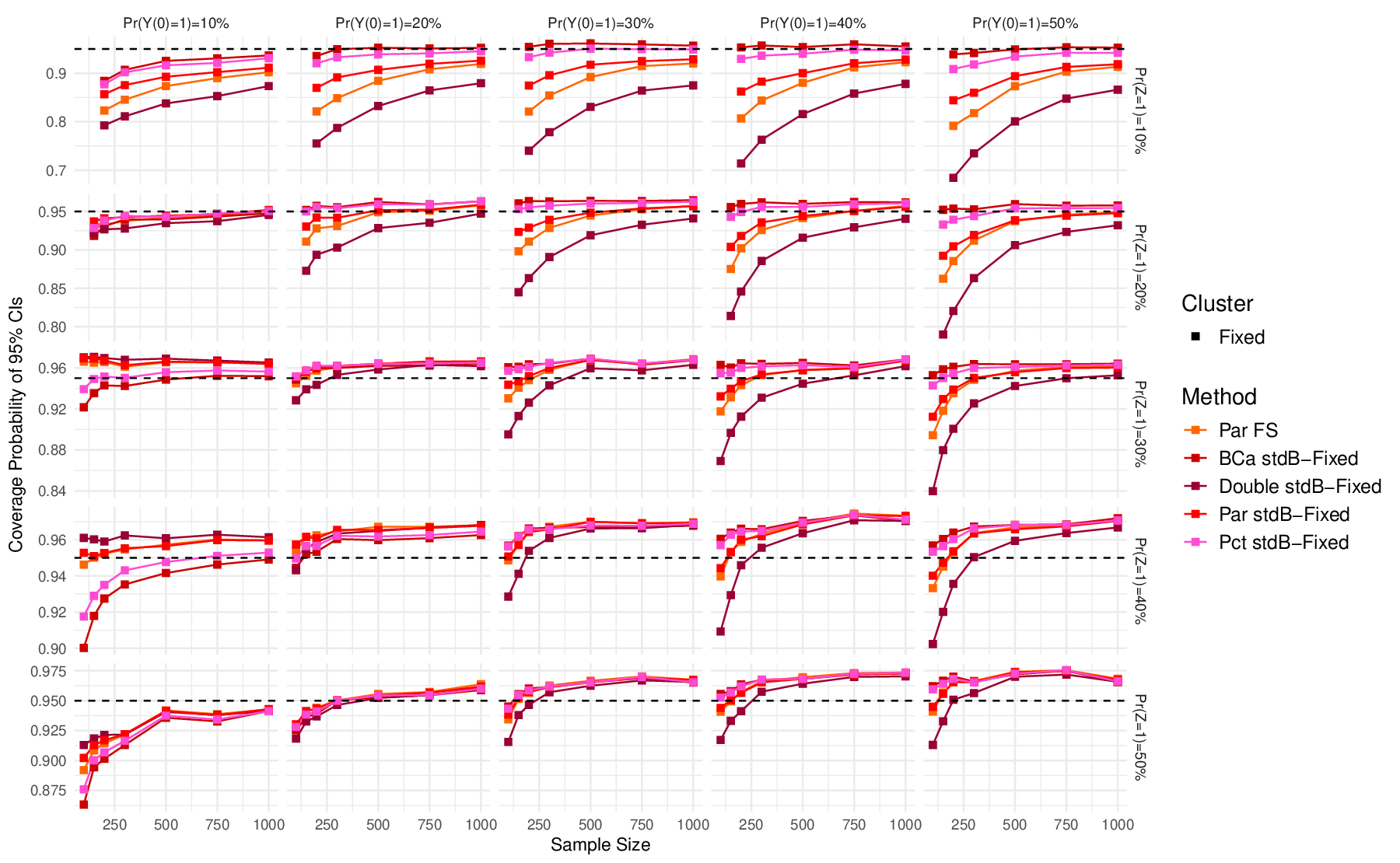}
    \caption{Empirical coverage rates of 95\% confidence intervals for the ATE: Results for Methods in Cluster \texttt{Fixed}}
    \label{fig: CProb_Cluster1}
\end{figure}

\begin{figure}[h]
    \centering    \includegraphics[scale = 0.55]{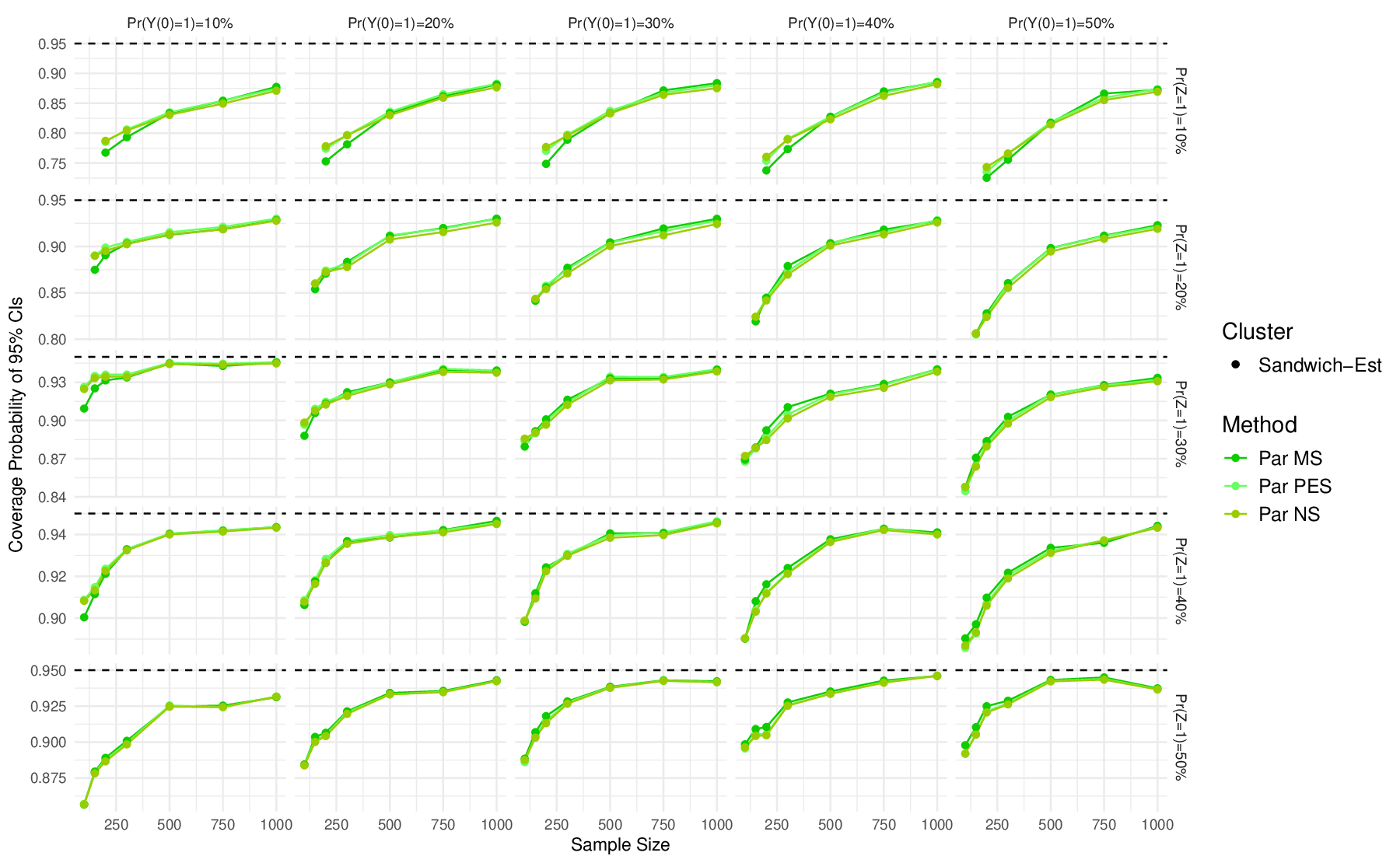}
    \caption{Empirical coverage rates of 95\% confidence intervals for the ATE: Results for Methods in Cluster \texttt{Sandwich-Est}}
    \label{fig: CProb_Cluster2}
\end{figure}

\begin{figure}[h]
    \centering    \includegraphics[scale = 0.55]{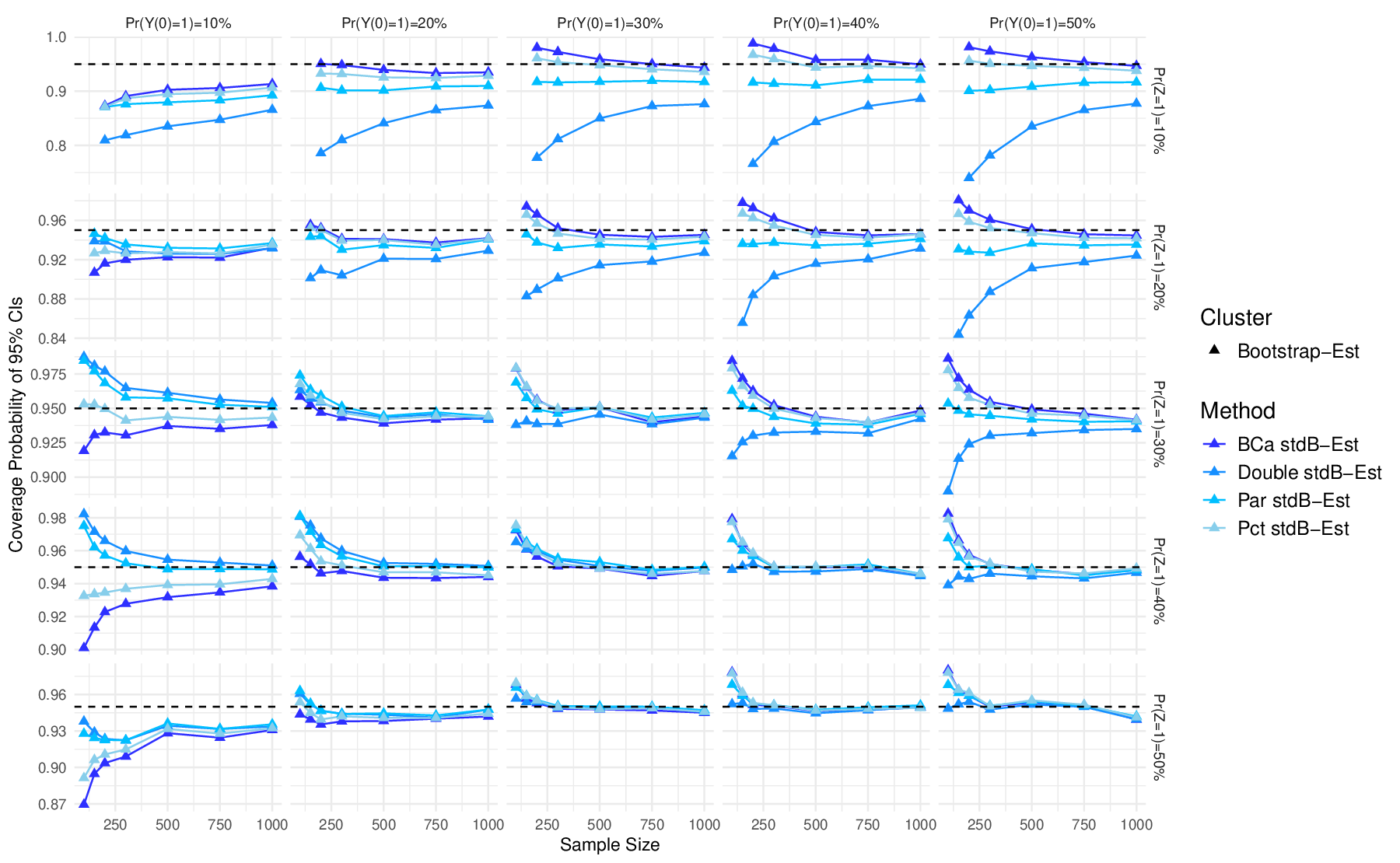}
    \caption{Empirical coverage rates of 95\% confidence intervals for the ATE: Results for Methods in Cluster \texttt{Bootstrap-Est}}
    \label{fig: CProb_Cluster3}
\end{figure}

\begin{figure}[h]
    \centering    \includegraphics[scale = 0.5]{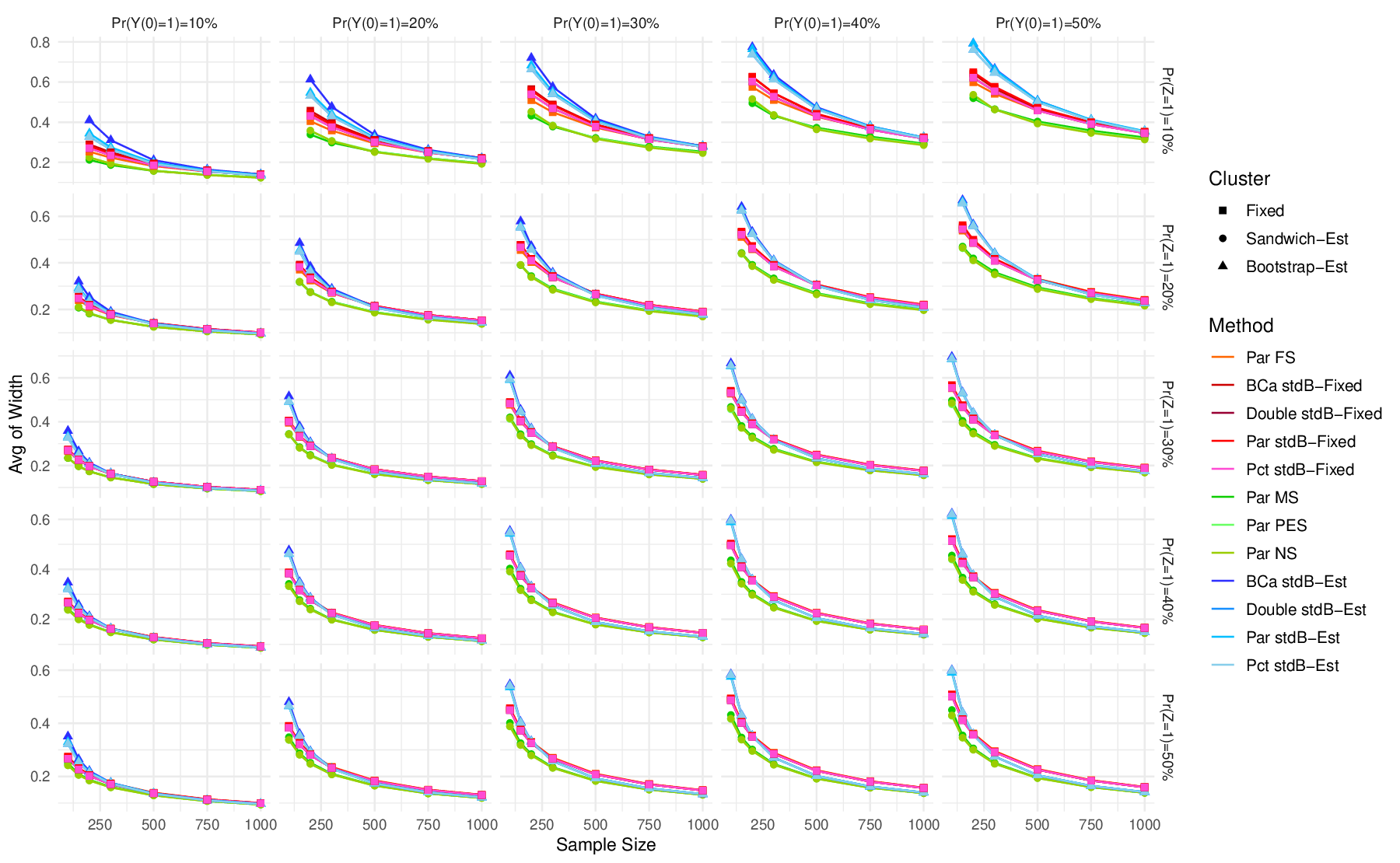}
    \caption{Average of Width for 95\% confidence intervals for the ATE: Results for all Methods}
    \label{fig: Width_Full}
\end{figure}

The Monte Carlo results for the empirical coverage rates of 95\% CIs are reported in Figure \ref{fig: CProb_Full}. To facilitate analysis, Figures \ref{fig: CProb_Cluster1},  Figures \ref{fig: CProb_Cluster2} and Figure \ref{fig: CProb_Cluster3} present these same results stratified by cluster. In addition, the average of width of these intervals are presents in Figure \ref{fig: Width_Full}.

\paragraph{Comparison of Confidence Interval Performance between Clusters:}
The coverage results are a direct consequence of the SE ratio behaviors. Generally speaking, across all scenarios for small sample sizes, methods from the \textbf{\texttt{Bootstrap-Est}} cluster yield the most conservative (i.e., highest coverage) results, often at or above the 95\% nominal level. This is followed by the \textbf{\texttt{Fixed}} cluster. In contrast, the \textbf{\texttt{Sandwich-Est}} cluster demonstrates poor coverage in small samples, falling well below 95\%, a direct result of its underestimation of the standard error.

As the sample size increases, the \textbf{\texttt{Bootstrap-Est}} methods converge correctly to 95\%, while the \textbf{\texttt{Fixed}} methods become overly conservative (with coverage exceeding 95\%), consistent with their SE over-estimation. The performance of the \textbf{\texttt{Sandwich-Est}} cluster improves with larger $n$, but its convergence heavily depends on the treatment allocation; larger samples are required to achieve nominal coverage when the allocation is highly imbalanced. The average CI widths (Figure \ref{fig: Width_Full}) further confirm this, showing the \textbf{\texttt{Sandwich-Est}} cluster always produces the narrowest CIs, and the \textbf{\texttt{Fixed}} cluster's CIs become the widest in large samples.


\subsubsection{Results for ATO}

\begin{figure}
    \centering    
    \includegraphics[scale = 0.5]{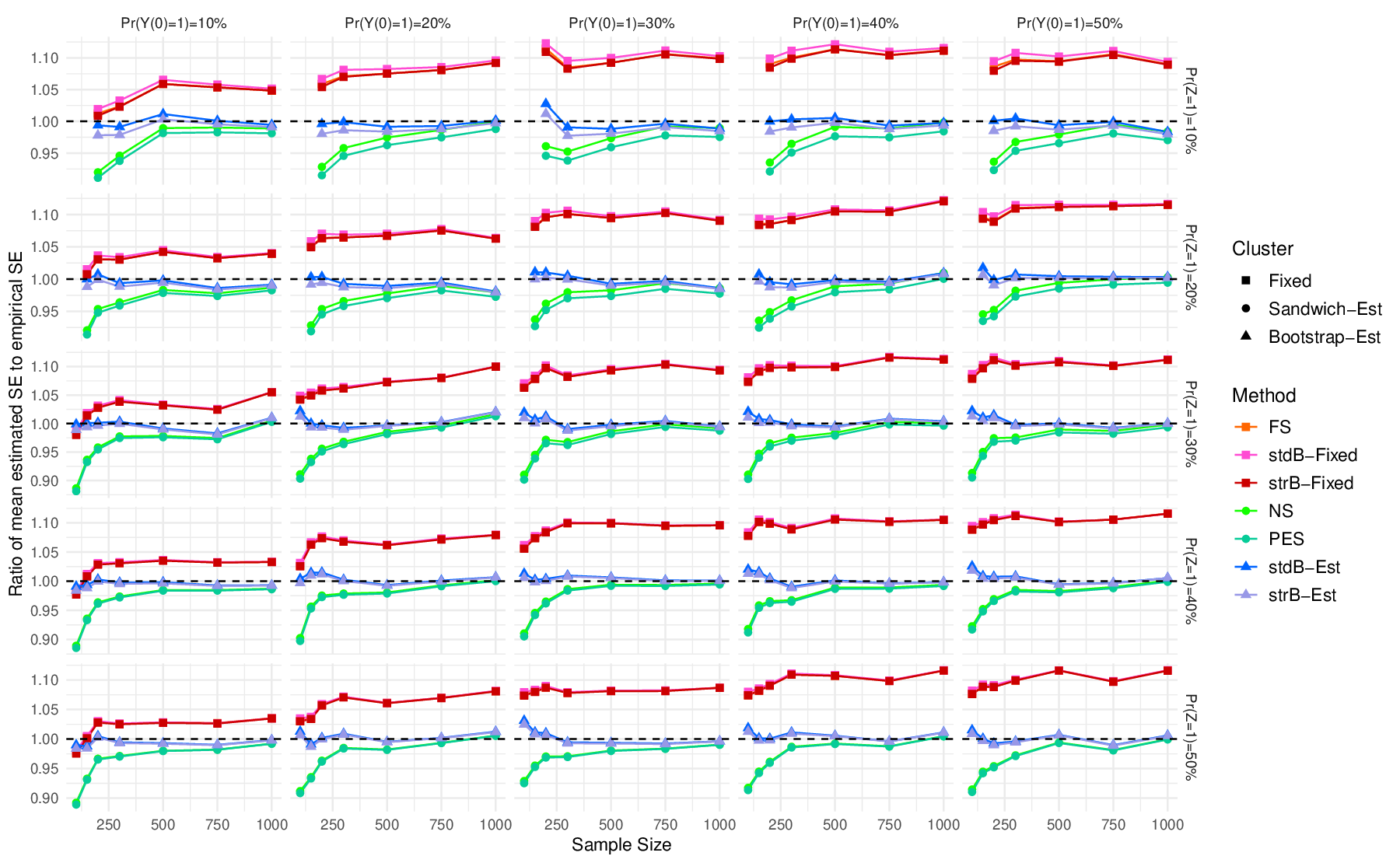}
    \captionsetup{justification=centering, singlelinecheck=off} % Center the caption and allow for footnote
    \caption[Caption for LOF]{Ratio of mean estimated SE\footnotemark to empirical SE for the ATO: Results for all Methods}
    \label{fig: SE.Ratio_Full_ATO}
\end{figure}

\begin{figure}
    \centering    \includegraphics[scale = 0.55]{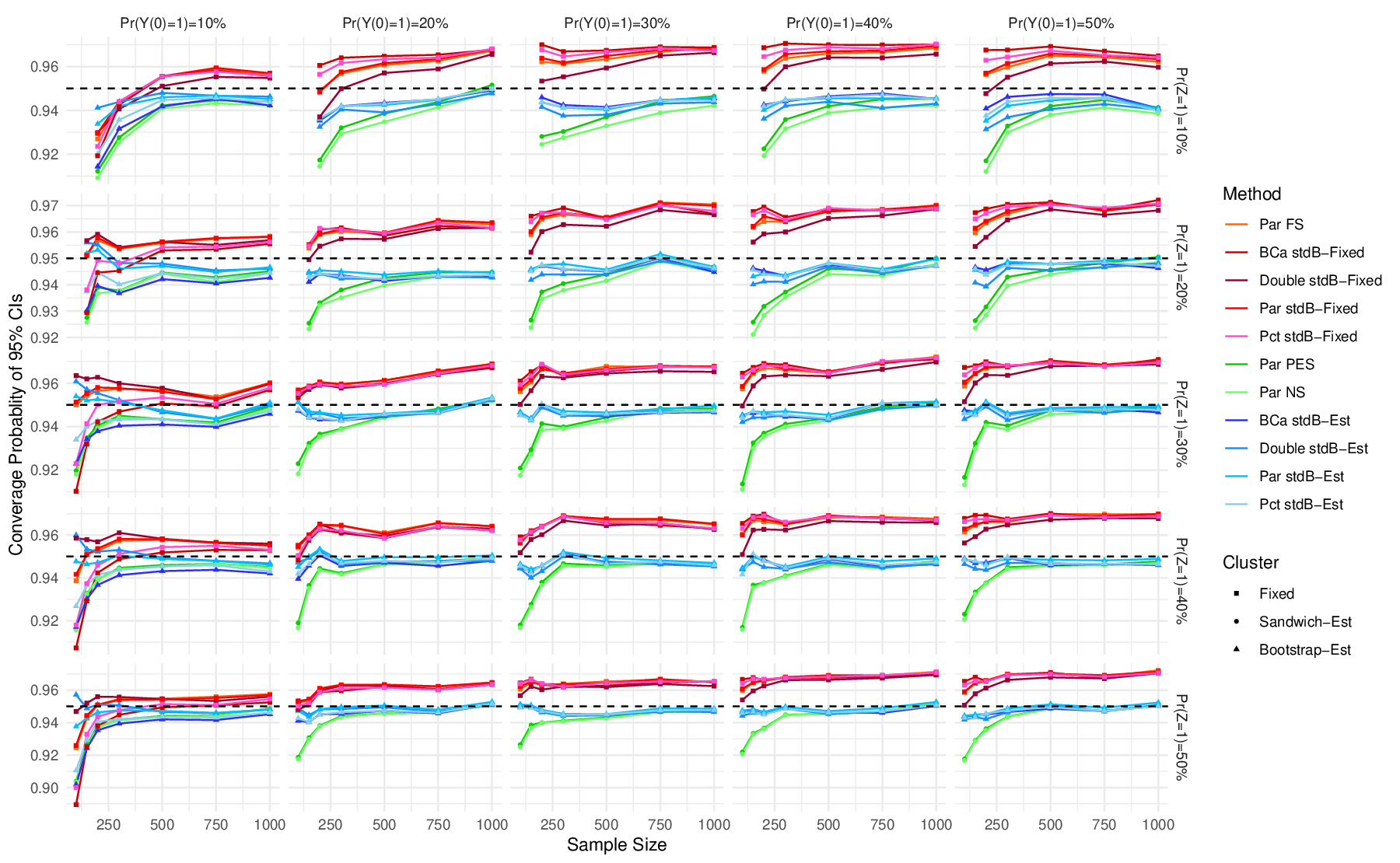}
    \caption{Empirical coverage rates of 95\% confidence intervals for the ATO: Results for all Methods}
    \label{fig: CProb_Full_ATO}
\end{figure}

We extended our analysis to the ATO to assess whether the same performance patterns and cluster structures persist. Results for SE estimation (Figure \ref{fig: SE.Ratio_Full_ATO}) and coverage probability (Figure \ref{fig: CProb_Full_ATO}) confirm that the three-cluster phenomenon is immediately apparent and behaves similarly to the ATE. However, these results also revealed a critical distinction: the estimators' asymptotic properties manifest more rapidly for the ATO, especially in small samples.

This enhanced stability is particularly evident in the most challenging scenarios. For instance, in cases of extreme treatment imbalance ($\Pr(Z=1)=10\%$), , all ATO methods provided more conservative and higher coverage rates than their ATE counterparts. This improved small-sample performance is most notable in the \textbf{\texttt{Sandwich-Est}} cluster; these methods, which performed poorly for the ATE, now achieve or approach the nominal 95\% coverage much more quickly for the ATO (e.g., with $n$ as low as 500 in some scenarios). Despite this significant improvement, this group remains the least conservative cluster in small samples.

In contrast, the \textbf{\texttt{Bootstrap-Est}} cluster remains robustly at or near the 95\% nominal level across all settings, consistently providing the most accurate coverage. The \textbf{\texttt{Fixed}} cluster continues to exhibit its characteristic inversion, becoming conservative in even smaller samples and becoming overly conservative with larger $n$. These findings suggest that ATO estimators, by down-weighting individuals with extreme propensity scores, are inherently more stable and less reliant on large sample sizes to achieve their target performance.

\footnotetext{The SEs of the mean estimated SE for 10,000 runs of the Monte Carlo simulation are all less than 0.001 across all methods, demonstrating the precision of our estimated SE results.}

\begin{figure}
    \centering    \includegraphics[scale = 0.5]{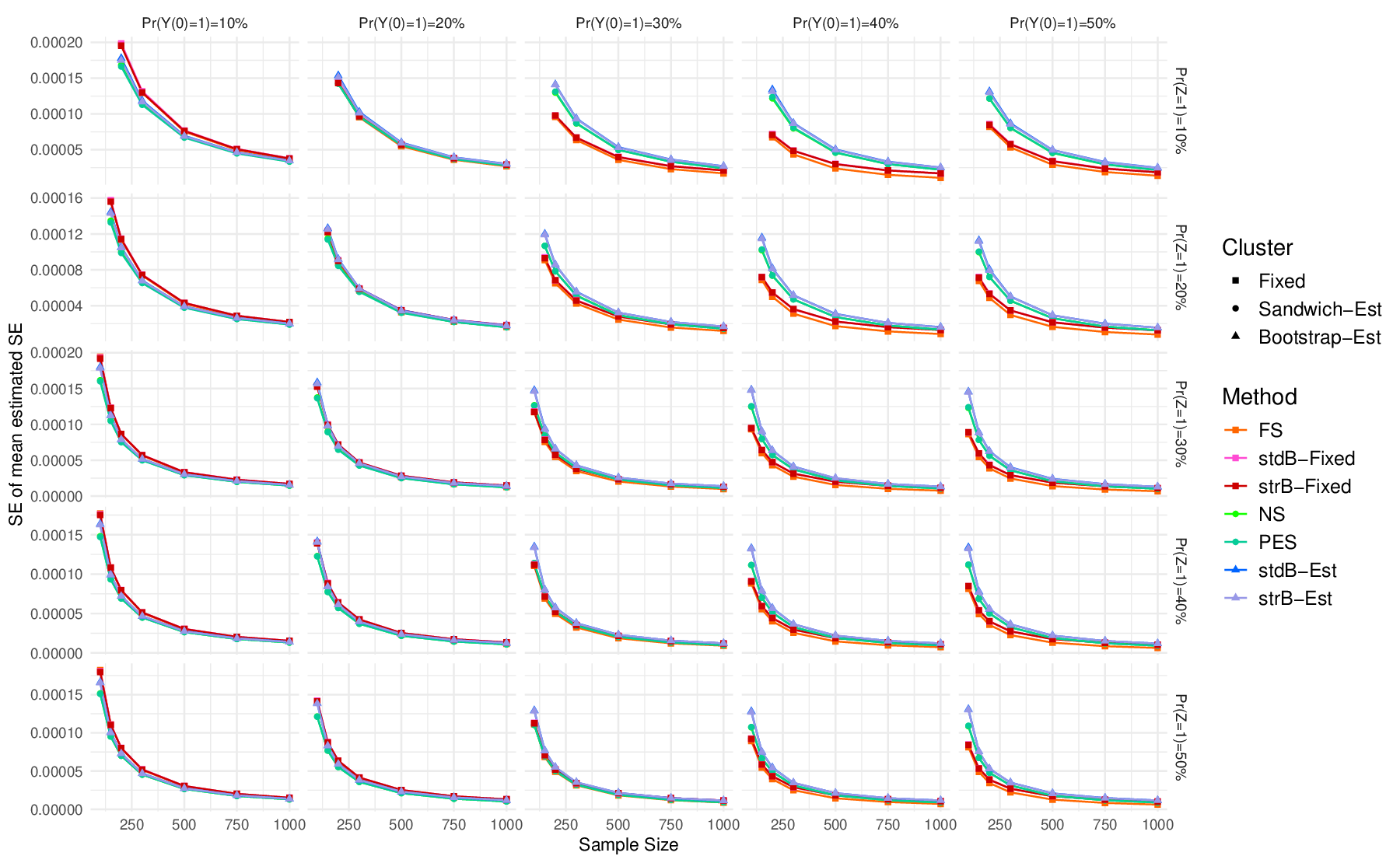}
    \caption{Standard error of all estimated SEs among Monte Carlo Simulation for ATO: Results for all methods}
    \label{fig: SE.SE_Full_ATO}
\end{figure}

\begin{figure}
    \centering    \includegraphics[scale = 0.5]{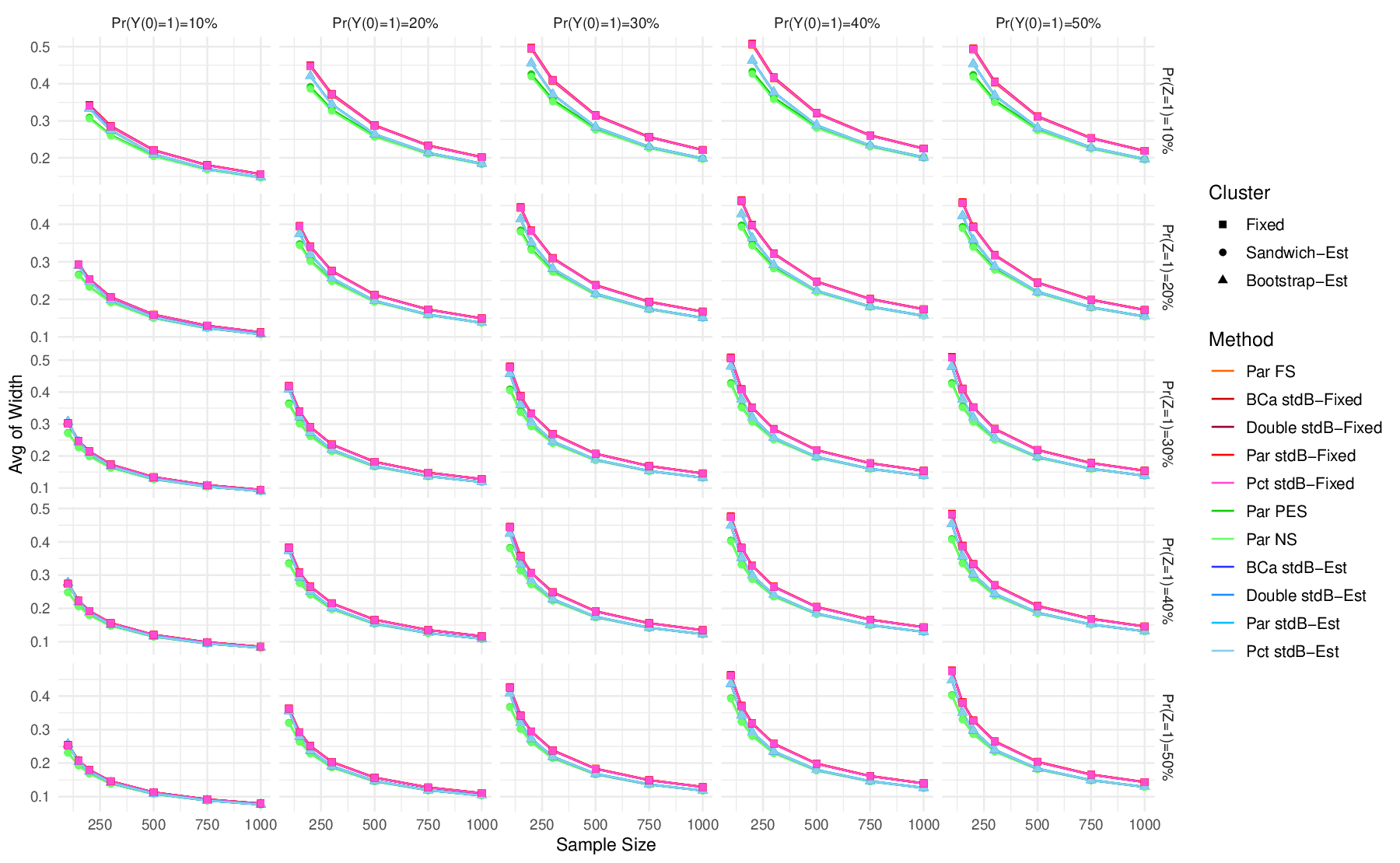}
    \caption{Average of Width for 95\% confidence intervals for the ATO: Results for all Methods}
    \label{fig: Width_Full_ATO}
\end{figure}

The remaining figures for the standard error of the SE (Figure \ref{fig: SE.SE_Full_ATO}) and CI Width (Figure \ref{fig: Width_Full_ATO}) further support these conclusions, confirming the persistence of the three-cluster structure and the enhanced small-sample stability of the ATO.

\end{document}